\documentclass[prl,twocolumn,amsmath,amssymb]{revtex4-1}

\usepackage{graphicx}
\usepackage{bm}
\usepackage{gensymb}
\usepackage{multirow}
\usepackage{xcolor}
\usepackage{csquotes}
\usepackage{braket}
\usepackage{amsmath}
\usepackage{hyperref} 
\newcommand{\beginsupplement}{%
        \setcounter{table}{0}
        \renewcommand{\thetable}{S\arabic{table}}%
        \setcounter{figure}{0}
        \renewcommand{\thefigure}{S\arabic{figure}}%
     }

\DeclareGraphicsExtensions{.eps , .jpg,.pdf,.png}

\begin{document}

\title{Kondo exhaustion and conductive surface states in antiferromagnetic YbIr$_3$Si$_7$}
	
\author{Macy Stavinoha$^1$, C.-L. Huang$^2$, W. Adam Phelan$^3$, Alannah M. Hallas$^2$, V. Loganathan$^2$, Jeffrey W. Lynn$^4$, Qingzhen Huang$^4$, Franziska Weickert$^5$, Vivien Zapf$^5$, Katharine R. Larsen$^6$, Patricia D. Sparks$^6$, James C. Eckert$^6$, Anand B. Puthirath$^7$, C. Hooley$^8$, Andriy H. Nevidomskyy$^2$, and E. Morosan$^{1,2}$}
	
\affiliation{$^1$Department of Chemistry, Rice University, Houston, TX 77005 USA
\\$^2$Department of Physics and Astronomy, Rice University, Houston, TX 77005 USA
\\$^3$Platform for the Accelerated Realization, Analysis, and Discovery of Interface Materials (PARADIM), Department of Chemistry, The Johns Hopkins University, Baltimore, MD 21218, USA
\\$^4$NIST Center for Neutron Research, National Institute of Standards and Technology, Gaithersburg, Maryland 20899, USA
\\$^5$National High Magnetic Field Laboratory, Materials Physics and Applications Division, Los Alamos
National Laboratory, Los Alamos, New Mexico 87545, USA
\\$^6$Department of Physics, Harvey Mudd College, Claremont, California 91711, USA
\\$^7$Department of Materials Science and NanoEngineering Rice University Houston, TX 77005, USA
\\$^8$Scottish Universities Physics Alliance, School of Physics and Astronomy, University of St Andrews, North Haugh, St. Andrews, Fife KY16 9SS, United Kingdom}
\date{\today}

	\maketitle

{\bf The interplay of Kondo screening and magnetic ordering in strongly correlated materials containing local moments is a subtle problem.\cite{schaffer2016} Usually the number of conduction electrons matches or exceeds the number of moments, and a Kondo-screened heavy Fermi liquid develops at low temperatures.\cite{kontani2004} Changing the pressure, magnetic field, or chemical doping can displace this heavy Fermi liquid in favor of a magnetically ordered state.\cite{stewart2001,stewart2006} Here we report the discovery of a version of such a `Kondo lattice' material, YbIr$_3$Si$_7$, in which the number of free charge carriers is much less than the number of local moments. This leads to `Kondo exhaustion':\cite{meyer2000} the electrical conductivity tends to zero at low temperatures as all the free carriers are consumed in the formation of Kondo singlets. This effect coexists with antiferromagnetic long-range order, with a N{\'e}el temperature $T\rm_N = 4.1\,{\rm K}$. Furthermore, the material shows conductive surface states with potential topological nature, and thus presents an exciting topic for future investigations.}

Kondo insulators represent a complex amalgam of competing interactions: Kondo screening, Ruderman-Kittel-Kasuya-Yosida (RKKY) interactions between nearby local moments, and crystal electric field effects \cite{dzero2016}. When strong spin-orbit coupling is factored in, non-trivial topological states may occur \cite{dzero2010}. Very few of the suggested Kondo insulator compounds show long-range magnetic order. Of those that do order magnetically, UFe$_4$P$_{12}$ is ferromagnetic below $3.1\,{\rm K}$ \cite{nakotte1999}, and CeOs$_2$Al$_{10}$ is antiferromagnetic below $28.5\,{\rm K}$ \cite{kawabata2015}. Here, we report the discovery of a new material, YbIr$_3$Si$_7$, which appears to be at the border of Kondo insulating behavior, and which also exhibits both long-range antiferromagnetic order and conductive surface states. The presence of surface states in YbIr$_3$Si$_7$ but not in YbRh$_3$Si$_7$ \cite{raiYbRh3Si7} favors a topology scenario, given the larger spin-orbit coupling in the former compound. In addition, it appears that the smaller chemical pressure at the surface induces a valence change from magnetic Yb$^{3+}$ in the bulk to the larger, non-magnetic Yb$^{2+}$ on the surface. Hence, the more a conductive surface could be a consequence of the lack of Kondo effect from the non-magnetic Yb$^{2+}$ ions. Conductive surface associated with a topological crystalline insulator state is also a possibility, albeit much less likely, given the low symmetry of the surface of the rhombohedral YbIr$_3$Si$_7$-type structure.\cite{raiYbRh3Si7} 

YbIr$_3$Si$_7$ exhibits several striking and unusual features. First, the usual mild Kondo upturn in the resistivity below the single-ion Kondo temperature $T\rm_K$ appears instead as an increase of the resistivity by several orders of magnitude. As we show below, the bulk part of the conductivity tends to zero in the low-temperature limit. Second, the onset of antiferromagnetic order does not prevent the continued rise of the resistivity even for $T < T\rm_N$. Third, this all happens in parallel with a much more gentle evolution of the surface transport, which remains conducting even at the lowest temperatures. We interpret the robust loss of carriers in terms of the Kondo exhaustion scenario, first proposed by Nozi{\`e}res \cite{Noz1,Noz2} and subsequently investigated numerically by Meyer and Nolting \cite{meyer2000}. The origin and nature of the surface states is less clear, but would merit further investigation, especially due to the potential for topological protection, hinted at by the strong spin-orbit coupling in this material, and by the lack of comparable surface conductivity in the isostructural compound YbRh$_3$Si$_7$ \cite{raiYbRh3Si7}.

\begin{figure*}[t!]
	\includegraphics[width=2\columnwidth]{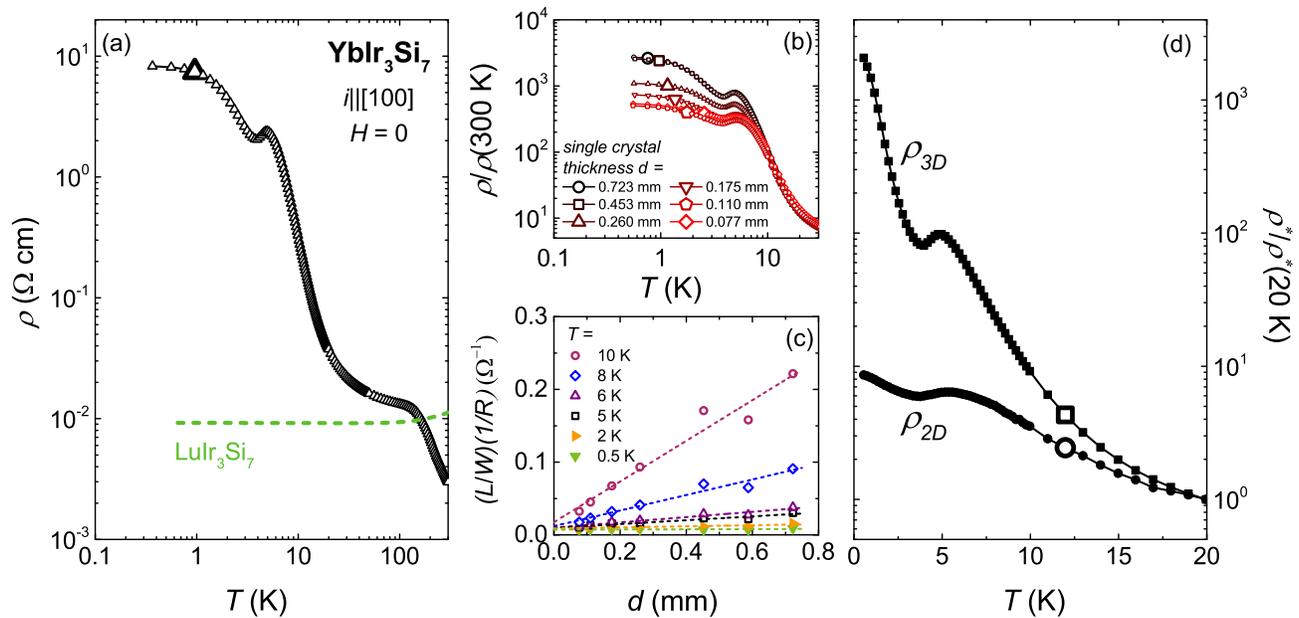}
	\caption{\label{Transport} (a) $H$ = 0 temperature-dependent resistivity for YbIr$_3$Si$_7$ single crystals (open triangles) and the polycrystalline non-magnetic analog LuIr$_3$Si$_7$ (dashed line). (b) $H$ = 0 scaled temperature-dependent $\rho(T)$ for different thickness values $d$ of a thinned crystal. (c) Inverse total resistance 1/$R$, scaled by the geometric factor $L$/$W$, as a function of $d$ for different temperatures (see text). (d-e) Scaled temperature-dependent resistivity $\rho^*/\rho^*(20 K)$ where $\rho^*$ is either the surface resistivity $\rho_{2D}$ (circles) or bulk resistivity $\rho_{3D}$ (squares) determined from the linear fits in (c).}
\end{figure*}

The physics of YbIr$_3$Si$_7$ appears to be distinct from both of the isostructural materials YbIr$_3$Ge$_7$ and YbRh$_3$Si$_7$ recently reported \cite{raiYbIr3Ge7,raiYbRh3Si7}, and from the known Kondo insulators CeFe$_4$P$_{12}$, Ce$_3$Bi$_4$Pt$_3$ \cite{hundley1990}, CeFe$_2$Al$_{10}$, CeRu$_4$Sn$_6$ \cite{das1995}, SmB$_6$ \cite{menth1969} and YbB$_{12}$ \cite{Kasaya1983,Kasaya1985}. YbIr$_3$Ge$_7$ and YbRh$_3$Si$_7$ show Kondo screening giving way to long-range magnetic order and are low-carrier or bad metals at low temperatures. Conversely, the other Kondo insulators mentioned above are paramagnets or intermediate-valence materials without long-range magnetic order. Long-range antiferro- or ferromagnetic order occurs only in CeOs$_2$Al$_{10}$ \cite{kawabata2015} and UFe$_4$P$_{12}$ \cite{meisner1985}, with YbIr$_3$Si$_7$ the first Yb-based material in the insulator class (albeit at the boundary of that class) that orders antiferromagnetically below $T\rm_N = 4.1\,{\rm K}$.

Fig.~\ref{Transport}(a) shows evidence for the insulating behavior in the $H$ = 0 electrical resistivity $\rho(T)$ of YbIr$_3$Si$_7$ (symbols). The data for the non-magnetic analog LuIr$_3$Si$_7$ (dashed line), where the $\rho(T)$ is weakly $T$-dependent and overall very large ($\sim10^{-2}$ $\Omega$ cm), demonstrate the low-carrier density nature of the $R$Ir$_3$Si$_7$ ($R$ = Yb, Lu) material class. We interpret the differences between the YbIr$_3$Si$_7$ and LuIr$_3$Si$_7$ curves as arising from the interaction between this low-carrier density conduction sea and the local moments of the holes in the Yb 4$f$ orbitals. A striking feature is that the YbIr$_3$Si$_7$ resistivity increases by $\sim$ 4 orders of magnitude on cooling from 300 K to 0.3 K. At low $T$, the resistivity levels off in a manner reminiscent of the plateau caused by conductive surface states in SmB$_6$ \cite{syers2015}. On the way to the low temperature plateau, two notable inflections occur in $\rho(T)$ for YbIr$_3$Si$_7$, above 100 K and below 10 K: the high temperature inflection is likely not related to long-range magnetic order, as it does not move in temperature between the $\mu_0H$ = 0 and $\mu_0H$ = 9 T (Fig. \ref{FigS2} in Supplementary Materials). Crystal electric field signatures in $\rho(T)$ can be ruled out as the origin of the 100 K feature, since anisotropic magnetization measurements (Fig. \ref{FigS4} in Supplementary Materials) point to strong crystal electric field effects at much larger temperatures ($>$ 400 K), with no visible features near 100 K. Furthermore, this feature in $\rho(T)$ disappears in the dilute moment limit, as we demonstrate with resistivity measurements on Yb$_{0.05}$Lu$_{0.95}$Ir$_3$Si$_7$ (Fig. \ref{FigS3} in Supplementary Materials), ruling out single-ion crystal electric field effects. Lastly, a structural phase transition can also be ruled out since neutron diffraction data at T = 1.5 K, shown below, confirm the room temperature structure. The most likely explanation for the 100 K resistivity feature is the onset of Kondo correlations around this temperature.


The description of the low-$T$ resistivity in YbIr$_3$Si$_7$ goes beyond the purely paramagnetic Kondo screening scenario. As neutron scattering and thermodynamic measurements reveal --- see below --- antiferromagnetic long-range order sets in at $T\rm_N$ = 4.1 K. The onset of this order is associated with a drop in resistivity for temperatures just below the N{\'e}el temperature, which we interpret as the temporary `liberation' of free carriers as the magnetic moments emerge from their screening clouds to form the N{\'e}el state. However, contrary to the usual behavior for $T \ll T\rm_N$, the resistivity then resumes its rise, suggesting that there are not enough carriers left to screen the remaining non-ordered fraction of each local moment.


The low-temperature resistivity of YbIr$_3$Si$_7$ is akin to the low-temperature plateau in the resistivity of SmB$_6$ attributed to conductive surface states \cite{phelan2014}. One test for the validity of this hypothesis is the dependence of the total resistivity on sample thickness $d$; the relative sample surface (conductive) to bulk volume (insulating) ratio increases when the thickness decreases, and should result in an overall resistivity decrease.  Fig. \ref{Transport}b confirms the low-$T$ resistivity decrease as the crystal is thinned from 723 $\mu$m to 77 $\mu$m. The insulating bulk dominates the resistivity above 400 $\mu$m (black), but for thinner samples the low-$T$ resistivity drops by an order of magnitude. For a more quantitative estimate of the contributions of the bulk resistance  $R_{3D}$ = $\rho_{3D}\frac{L}{Wd}$ and surface resistance $R_{2D}$ = $\rho_{2D}\frac{L}{W}$ to the total resistance $R$, ($L$ = voltage drop distance, $W$ = sample width, $d$ = sample thickness), one can write

\begin{equation}
\frac{1}{R} = \frac{1}{R_{3D}} + \frac{1}{R_{2D}} = \frac{W}{\rho_{3D}L}d + \frac{W}{\rho_{2D}L} \nonumber \\
\end{equation}
or
\begin{equation}
\frac{1}{R} = \frac{W}{L}\left(\frac{1}{\rho_{3D}}d + \frac{1}{\rho_{2D}}\right) \nonumber \\ 
\end{equation}
which gives
\begin{equation}
\frac{L}{W}\frac{1}{R} = \frac{1}{\rho_{3D}}d + \frac{1}{\rho_{2D}} \label{equation}\\
\end{equation}

\begin{figure}[b!]
	\includegraphics[width=1\columnwidth]{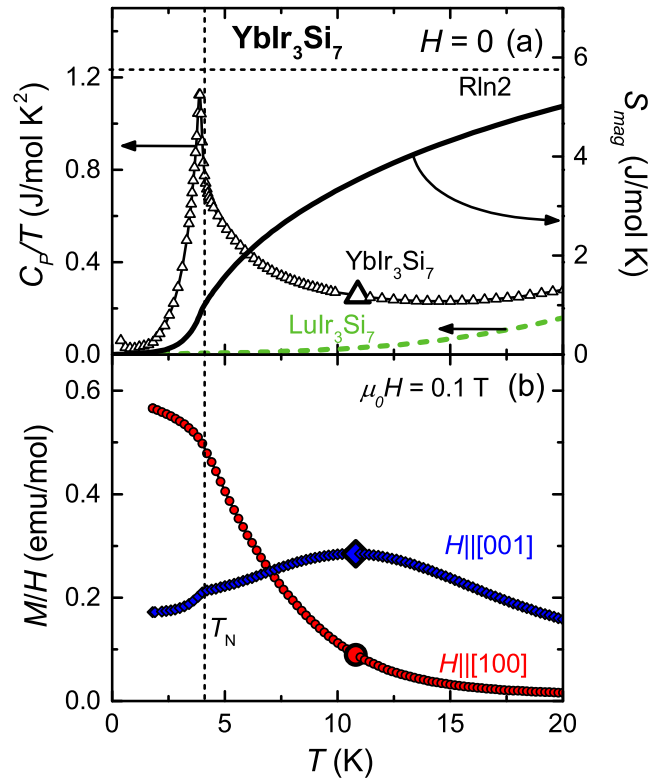}
	\caption{\label{Mag} (a) $H$ = 0 specific heat scaled by temperature $C_P/T$ (left axis) for YbIr$_3$Si$_7$ (symbols) and LuIr$_3$Si$_7$ (dashed line). Right axis: YbIr$_3$Si$_7$ magnetic entropy $S_{mag}$ (solid line). (b) Low temperature magnetic susceptibility $M$/$H$ for $\mu_0H$ = 0.1 T along $H$$\parallel$[100] (red circles) and  $H$$\parallel$[001] (blue diamonds). Vertical dashed line marks the antiferromagnetic ordering temperature $T\rm_N$.}
\end{figure}

When plotting $\frac{L}{W}\frac{1}{R}$ as a function of thickness $d$ (Fig. \ref{Transport}c), eqn. \ref{equation} suggests that different isothermal plots should be lines with the slope and intercept inversely proportional to $\rho_{3D}$ and $\rho_{2D}$, respectively. This is true \textit{if} the two contributions $\rho_{3D}$ and $\rho_{2D}$ are thickness-independent, and such an assumption is indeed validated by the linear plots shown in Fig. \ref{Transport}c (symbols). The dotted lines are fits to eqn. \ref{equation}, from which the $\rho_{3D}(T)$ and $\rho_{2D}(T)$ values are extracted. A comparison between the surface and bulk resistivities, scaled at 20 K (Fig. \ref{Transport}d) reveals a much weaker temperature dependence of $\rho_{2D}$ compared to $\rho_{3D}$, consistent with a much more conductive surface compared to the bulk.

Thermodynamic measurements reveal long-range magnetic order in YbIr$_3$Si$_7$ below $T\rm_N$ = 4.1 K (Fig \ref{Mag}). The zero-field specific heat $C_P/T$ data for YbIr$_3$Si$_7$ (symbols, Fig \ref{Mag}a) show a peak at 4.1 K, while the non-magnetic analog LuIr$_3$Si$_7$ (dashed line) displays much smaller $C_P/T$ values and no phase transition. Above the ordering temperature, a large tail in the YbIr$_3$Si$_7$  specific heat could be the result of short-range magnetic interactions. The magnetic entropy $S_{mag}$ for YbIr$_3$Si$_7$ (solid line, Fig. \ref{Mag}a) was estimated from $S_{mag}$ =$\int_{0}^{T} \frac{C_{mag}}{T}~dT$, where $C_{mag} = C_P$(YbIr$_{3}$Si$_{7}) - C_P$(LuIr$_{3}$Si$_{7}$). At $T\rm_N$, the low entropy release of $\sim$ 15\% $R$ln2 signals Kondo screening, with an estimated Kondo temperature $T\rm_K~\approx$ 16 K from $S_{mag}(0.5\,T_{\mathrm{K}})=0.5~R\,\mathrm{ln}2$. With this $T\rm_K$ estimate in mind, the low temperature inflection in the $H$ = 0 YbIr$_3$Si$_7$ resistivity data (Fig. \ref{Transport}a) can be correlated to Kondo screening.

Further confirmation of the long-range magnetic order is provided through anisotropic magnetic susceptibility measurements $M$/$H$ shown in Fig. \ref{Mag}. Temperature-dependent $M$/$H$ measurements with $\mu_0H$ = 0.1 T confirm the magnetic phase transition as small inflections are observed on cooling for both $H\parallel$[100] (circles) and $H\parallel$[001] (diamonds). On warming above $T\rm_N$, the $H\parallel$[001] data (full diamonds in Fig. \ref{Mag}b) display a broad maximum centered around 10 K, while a similar feature at 10 K is absent in the [100] field orientation. The crossing of the anisotropic $M$/$H$ curves just above $T\rm_N$ is reminiscent of the hard axis ordering observed in both known isostructural Yb systems YbIr$_3$Ge$_7$\cite{raiYbIr3Ge7}, YbRh$_3$Si$_7$\cite{raiYbRh3Si7}, \textit{and} in the few known FM Kondo lattice compounds.\cite{hafner2019} Indeed, the susceptibility at high temperatures (Fig. \ref{FigS4} in Supplementary Materials) is indicative of large crystal electric field anisotropy favoring a [001] easy axis. The average magnetization $M_{ave}$ was calculated using $M_{ave}$ = $\frac{1}{3}$(2$M_{100}$~+~$M_{001}$). A linear fit to the average inverse susceptibility at high temperatures (above 350 K) yields an effective moment $\mu_{eff}$ = 4.42 $\mu_B$, which is close to the theoretical value of 4.54 $\mu_B$ for Yb$^{3+}$.

\begin{figure}[t!]
	\includegraphics[width=1\columnwidth]{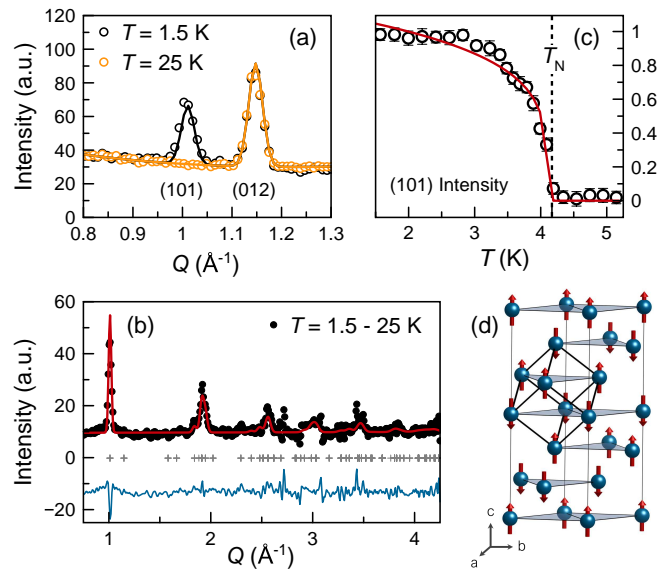}
	\caption{(a) Powder neutron diffraction measurements for YbIr$_3$Si$_7$ at $T = 25$~K (orange symbols) and $T = 1.5$~K (black symbols). (b) Rietveld refinement (red line) with the $k=0$ $\Gamma_1$ irreducible representation of the magnetic diffraction pattern (black symbols), which is obtained by subtracting the $T = 25$ K from the $T = 1.5$ K data. The difference between experiment and refinement is given by the blue line and the symmetry-allowed peak positions for the $R\bar{3}c$ space group are given by the crosses. (c) Neutron order parameter from the intensity of the (101) magnetic Bragg peak. (d) The spin configuration of the Yb$^{3+}$ moments in the $\Gamma_1$ ordered state.}
\label{Neutron} 
\end{figure}

In order to elucidate the nature of the magnetic ground state, we performed powder neutron diffraction measurements. Upon cooling from 25 K (orange symbols, Fig. \ref{Neutron}a) to 1.5 K (black symbols, Fig. \ref{Neutron}a), we observe the formation of magnetic Bragg peaks, which can be indexed with the propagation vector $k~=~0$. There are three possible $k~=~0$ irreducible representations, $\Gamma_1$, $\Gamma_3$, and $\Gamma_5$, for Yb$^{3+}$ sitting at the $6b$ Wyckoff site within the $R\bar{3}c$ space group. The latter two of these possible magnetic structures can be ruled out, since they would generate intense magnetic Bragg peaks at positions where no peaks are observed in experiment. We are left to consider only the $\Gamma_1$ irreducible representation, and the Rietveld refinement with this magnetic structure is shown in Fig. \ref{Neutron}b. The magnetic diffraction pattern was isolated by subtracting a data set collected at $T=25$~K from a data set collected at $T = 1.5$~K. This gives excellent agreement between the refinement (red line in Fig. \ref{Neutron}b) and the measured data (black symbols). Furthermore, the order parameter (Fig. \ref{Neutron}c), which is derived by fitting the intensity of the (101) magnetic Bragg peak, drops to 0 at $T\rm_N$ = 4.1 K, consistent with the thermodynamic data (Fig. \ref{Mag}). In the $\Gamma_1$ antiferromagnetically-ordered state (Fig. \ref{Neutron}d), all the Yb$^{3+}$ moments are oriented along the crystallographic $c$ axis ([001]). Each Yb$^{3+}$ moment is aligned anti-parallel with its six nearest neighbors in the nearly cubic Yb sublattice (outlined in Fig. \ref{Neutron}d), and parallel with its co-planar next nearest neighbors (grey sheets). The refined ordered moment is 1.51(5) $\mu_B$/Yb$^{3+}$. 

The magnetization (Fig. \ref{Mag}) and neutron diffraction data (Fig. \ref{Neutron}) on YbIr$_3$Si$_7$ confirm the trivalent state of the Yb ions in the bulk. However, XPS measurements probe a smaller volume-to-surface ratio than bulk measurements, and are more sensitive to surface valence changes. Our XPS data (Supplementary Materials, Fig. \ref{FigS8}) on YbIr$_3$Si$_7$ reveal mixed valence of the Yb ions in the measured volume close to the surface, suggesting a valence change from Yb$^{3+}$ (smaller) in the bulk to Yb$^{2+}$ (larger) on the surface. This is possibly due to the smaller chemical pressure at the crystal surface,\cite{johansson1979} and, in turn, may be responsible for the enhanced surface conductivity in the absence of Kondo effect from the non-magnetic Yb$^{2+}$ ions.

YbIr$_3$Si$_7$ shows a remarkable portfolio of physical properties that makes it unique among known Kondo insulators: the Kondo insulating behavior is accompanied by long range AFM order, large crystal electric field anisotropy and conductive surface states. This provides a complex energy landscape to explore a variety of quantum interactions and their interplay with surface topology. To determine the cause of the insulating behavior, we performed DFT+U calculations to map the band structure of YbIr$_3$Si$_7$ (see the Supplementary Materials). The band structure calculation, however, is unable to capture the opening of a gap and shows bands crossing the Fermi level even when accounting for strong Kondo hybridization, indicating that the insulating behavior is a many-body effect. To fully understand the arrangement of competing interactions responsible for the insulating behavior, more sophisticated methods such as DMFT calculations are warranted. Further experimental investigation on YbIr$_3$Si$_7$ will include ARPES and transport as a function of sample morphology (underway), as well as point contact spectroscopy, aimed at characterizing the origin of the surface and bulk properties. However, some of these experiments are impeded by the three-dimensional rhombohedral structure, which does not lend itself to cleaving as is needed for some of these experiments.

\textbf{Acknowledgments.} MS, CLH, and EM acknowledge support from the Gordon and Betty Moore Foundation EPiQS initiative through grant GBMF 4417. EM also acknowledges partial travel support to Max Planck Institute in Dresden, Germany from the Alexander von Humboldt Foundation Fellowship for Experienced Researchers, where part of the work was carried out. Research at PARADIM (Platform for the Accelerated Realization, Analysis, and Discovery of Interface Materials) was supported by the National Science Foundation grant no. DMR-1539918. VL and AHN were supported by the Robert A. Welch Foundation grant C-1818. AHN also acknowledges the support of the National Science Foundation grant no. DMR-1350237. A portion of this work was performed at the National High Magnetic Field Laboratory, which is supported by the National Science Foundation Cooperative Agreement No. DMR-1644779, the State of Florida, and the United States Department of Energy. The use of the EPMA facility in the Department of Earth Science, Rice University is kindly acknowledged. KRL, PDS, and JCE would like to acknowledge support of the Jean Perkins Foundation, Florida State University, and Los Alamos National Laboratory. ABP thanks the Indo-US Science and Technology Forum (IUSSTF) for financial support in the form of a postdoctoral fellowship. CAH gratefully acknowledges financial support from the Engineering and Physical Sciences Research Council (UK) via grant number EP/R031924/1. He also thanks Rice University for its hospitality during a four-month visiting professorship where some of this work was carried out. We are grateful for fruitful discussions with B. Rai, T. McQueen, N. Caroca-Canales, C. Geibel, P. Coleman and L. Balicas.

\textbf{Note:} While our manuscript has been under review at Nature Physics, Nakamura $\emph{et al}$. reported (\emph{J. Phys. Soc. Jpn}. \textbf{88}, 093705) the properties of YbIr$_3$Si$_7$ single crystals grown from Sb flux. The properties they report are drastically different from those on our single crystals, which, in our experience, is very likely associated with Sb incorporation.  

\bibliography{submission_arXiv}

\begin{thebibliography}{30}%
\makeatletter
\providecommand \@ifxundefined [1]{%
 \@ifx{#1\undefined}
}%
\providecommand \@ifnum [1]{%
 \ifnum #1\expandafter \@firstoftwo
 \else \expandafter \@secondoftwo
 \fi
}%
\providecommand \@ifx [1]{%
 \ifx #1\expandafter \@firstoftwo
 \else \expandafter \@secondoftwo
 \fi
}%
\providecommand \natexlab [1]{#1}%
\providecommand \enquote  [1]{``#1''}%
\providecommand \bibnamefont  [1]{#1}%
\providecommand \bibfnamefont [1]{#1}%
\providecommand \citenamefont [1]{#1}%
\providecommand \href@noop [0]{\@secondoftwo}%
\providecommand \href [0]{\begingroup \@sanitize@url \@href}%
\providecommand \@href[1]{\@@startlink{#1}\@@href}%
\providecommand \@@href[1]{\endgroup#1\@@endlink}%
\providecommand \@sanitize@url [0]{\catcode `\\12\catcode `\$12\catcode
  `\&12\catcode `\#12\catcode `\^12\catcode `\_12\catcode `\%12\relax}%
\providecommand \@@startlink[1]{}%
\providecommand \@@endlink[0]{}%
\providecommand \url  [0]{\begingroup\@sanitize@url \@url }%
\providecommand \@url [1]{\endgroup\@href {#1}{\urlprefix }}%
\providecommand \urlprefix  [0]{URL }%
\providecommand \Eprint [0]{\href }%
\providecommand \doibase [0]{http://dx.doi.org/}%
\providecommand \selectlanguage [0]{\@gobble}%
\providecommand \bibinfo  [0]{\@secondoftwo}%
\providecommand \bibfield  [0]{\@secondoftwo}%
\providecommand \translation [1]{[#1]}%
\providecommand \BibitemOpen [0]{}%
\providecommand \bibitemStop [0]{}%
\providecommand \bibitemNoStop [0]{.\EOS\space}%
\providecommand \EOS [0]{\spacefactor3000\relax}%
\providecommand \BibitemShut  [1]{\csname bibitem#1\endcsname}%
\let\auto@bib@innerbib\@empty
\bibitem [{\citenamefont {Schaffer}\ \emph {et~al.}(2016)\citenamefont
  {Schaffer}, \citenamefont {Kin-Ho~Lee}, \citenamefont {Yang},\ and\
  \citenamefont {Kim}}]{schaffer2016}%
  \BibitemOpen
  \bibfield  {author} {\bibinfo {author} {\bibfnamefont {R.}~\bibnamefont
  {Schaffer}}, \bibinfo {author} {\bibfnamefont {E.}~\bibnamefont
  {Kin-Ho~Lee}}, \bibinfo {author} {\bibfnamefont {B.-J.}\ \bibnamefont
  {Yang}}, \ and\ \bibinfo {author} {\bibfnamefont {Y.~B.}\ \bibnamefont
  {Kim}},\ }\href@noop {} {\bibfield  {journal} {\bibinfo  {journal} {Reports
  on Progress in Physics}\ }\textbf {\bibinfo {volume} {79}},\ \bibinfo {pages}
  {094504} (\bibinfo {year} {2016})}\BibitemShut {NoStop}%
\bibitem [{\citenamefont {Kontani}\ and\ \citenamefont
  {Yamada}(2005)}]{kontani2004}%
  \BibitemOpen
  \bibfield  {author} {\bibinfo {author} {\bibfnamefont {H.}~\bibnamefont
  {Kontani}}\ and\ \bibinfo {author} {\bibfnamefont {K.}~\bibnamefont
  {Yamada}},\ }\href@noop {} {\bibfield  {journal} {\bibinfo  {journal}
  {Journal of the Physical Society of Japan}\ }\textbf {\bibinfo {volume}
  {74}},\ \bibinfo {pages} {155} (\bibinfo {year} {2005})}\BibitemShut
  {NoStop}%
\bibitem [{\citenamefont {Stewart}(2001)}]{stewart2001}%
  \BibitemOpen
  \bibfield  {author} {\bibinfo {author} {\bibfnamefont {G.~R.}\ \bibnamefont
  {Stewart}},\ }\href@noop {} {\bibfield  {journal} {\bibinfo  {journal}
  {Review of Modern Physics}\ }\textbf {\bibinfo {volume} {73}},\ \bibinfo
  {pages} {797} (\bibinfo {year} {2001})}\BibitemShut {NoStop}%
\bibitem [{\citenamefont {Stewart}(2006)}]{stewart2006}%
  \BibitemOpen
  \bibfield  {author} {\bibinfo {author} {\bibfnamefont {G.~R.}\ \bibnamefont
  {Stewart}},\ }\href@noop {} {\bibfield  {journal} {\bibinfo  {journal}
  {Review of Modern Physics}\ }\textbf {\bibinfo {volume} {78}},\ \bibinfo
  {pages} {743} (\bibinfo {year} {2006})}\BibitemShut {NoStop}%
\bibitem [{\citenamefont {Meyer}\ and\ \citenamefont
  {Nolting}(2000)}]{meyer2000}%
  \BibitemOpen
  \bibfield  {author} {\bibinfo {author} {\bibfnamefont {D.}~\bibnamefont
  {Meyer}}\ and\ \bibinfo {author} {\bibfnamefont {W.}~\bibnamefont
  {Nolting}},\ }\href@noop {} {\bibfield  {journal} {\bibinfo  {journal}
  {Physical Review B}\ }\textbf {\bibinfo {volume} {61}},\ \bibinfo {pages}
  {13465} (\bibinfo {year} {2000})}\BibitemShut {NoStop}%
\bibitem [{\citenamefont {Dzero}\ \emph {et~al.}(2016)\citenamefont {Dzero},
  \citenamefont {Xia}, \citenamefont {Galitski},\ and\ \citenamefont
  {Coleman}}]{dzero2016}%
  \BibitemOpen
  \bibfield  {author} {\bibinfo {author} {\bibfnamefont {M.}~\bibnamefont
  {Dzero}}, \bibinfo {author} {\bibfnamefont {J.}~\bibnamefont {Xia}}, \bibinfo
  {author} {\bibfnamefont {V.}~\bibnamefont {Galitski}}, \ and\ \bibinfo
  {author} {\bibfnamefont {P.}~\bibnamefont {Coleman}},\ }\href@noop {}
  {\bibfield  {journal} {\bibinfo  {journal} {Annual Review of Condensed Matter
  Physics}\ }\textbf {\bibinfo {volume} {7}},\ \bibinfo {pages} {249} (\bibinfo
  {year} {2016})}\BibitemShut {NoStop}%
\bibitem [{\citenamefont {Dzero}\ \emph {et~al.}(2010)\citenamefont {Dzero},
  \citenamefont {Sun}, \citenamefont {Galitski},\ and\ \citenamefont
  {Coleman}}]{dzero2010}%
  \BibitemOpen
  \bibfield  {author} {\bibinfo {author} {\bibfnamefont {M.}~\bibnamefont
  {Dzero}}, \bibinfo {author} {\bibfnamefont {K.}~\bibnamefont {Sun}}, \bibinfo
  {author} {\bibfnamefont {V.}~\bibnamefont {Galitski}}, \ and\ \bibinfo
  {author} {\bibfnamefont {P.}~\bibnamefont {Coleman}},\ }\href@noop {}
  {\bibfield  {journal} {\bibinfo  {journal} {Physical review letters}\
  }\textbf {\bibinfo {volume} {104}},\ \bibinfo {pages} {106408} (\bibinfo
  {year} {2010})}\BibitemShut {NoStop}%
\bibitem [{\citenamefont {Nakotte}\ \emph {et~al.}(1999)\citenamefont
  {Nakotte}, \citenamefont {Dilley}, \citenamefont {Torikachvili},
  \citenamefont {Bordallo}, \citenamefont {Maple}, \citenamefont {Chang},
  \citenamefont {Christianson}, \citenamefont {Schultz}, \citenamefont
  {Majkrzak},\ and\ \citenamefont {Shirane}}]{nakotte1999}%
  \BibitemOpen
  \bibfield  {author} {\bibinfo {author} {\bibfnamefont {H.}~\bibnamefont
  {Nakotte}}, \bibinfo {author} {\bibfnamefont {N.}~\bibnamefont {Dilley}},
  \bibinfo {author} {\bibfnamefont {M.}~\bibnamefont {Torikachvili}}, \bibinfo
  {author} {\bibfnamefont {H.}~\bibnamefont {Bordallo}}, \bibinfo {author}
  {\bibfnamefont {M.}~\bibnamefont {Maple}}, \bibinfo {author} {\bibfnamefont
  {S.}~\bibnamefont {Chang}}, \bibinfo {author} {\bibfnamefont
  {A.}~\bibnamefont {Christianson}}, \bibinfo {author} {\bibfnamefont
  {A.}~\bibnamefont {Schultz}}, \bibinfo {author} {\bibfnamefont
  {C.}~\bibnamefont {Majkrzak}}, \ and\ \bibinfo {author} {\bibfnamefont
  {G.}~\bibnamefont {Shirane}},\ }\href@noop {} {\bibfield  {journal} {\bibinfo
   {journal} {Physica B: Condensed Matter}\ }\textbf {\bibinfo {volume}
  {259}},\ \bibinfo {pages} {280} (\bibinfo {year} {1999})}\BibitemShut
  {NoStop}%
\bibitem [{\citenamefont {Kawabata}\ \emph {et~al.}(2015)\citenamefont
  {Kawabata}, \citenamefont {Ekino}, \citenamefont {Yamada}, \citenamefont
  {Sakai}, \citenamefont {Sugimoto}, \citenamefont {Muro},\ and\ \citenamefont
  {Takabatake}}]{kawabata2015}%
  \BibitemOpen
  \bibfield  {author} {\bibinfo {author} {\bibfnamefont {J.}~\bibnamefont
  {Kawabata}}, \bibinfo {author} {\bibfnamefont {T.}~\bibnamefont {Ekino}},
  \bibinfo {author} {\bibfnamefont {Y.}~\bibnamefont {Yamada}}, \bibinfo
  {author} {\bibfnamefont {Y.}~\bibnamefont {Sakai}}, \bibinfo {author}
  {\bibfnamefont {A.}~\bibnamefont {Sugimoto}}, \bibinfo {author}
  {\bibfnamefont {Y.}~\bibnamefont {Muro}}, \ and\ \bibinfo {author}
  {\bibfnamefont {T.}~\bibnamefont {Takabatake}},\ }\href@noop {} {\bibfield
  {journal} {\bibinfo  {journal} {Physical Review B}\ }\textbf {\bibinfo
  {volume} {92}},\ \bibinfo {pages} {201113} (\bibinfo {year}
  {2015})}\BibitemShut {NoStop}%
\bibitem [{\citenamefont {Rai}\ \emph {et~al.}(2018)\citenamefont {Rai},
  \citenamefont {Chikara}, \citenamefont {Ding}, \citenamefont {Oswald},
  \citenamefont {Sch\"onemann}, \citenamefont {Loganathan}, \citenamefont
  {Hallas}, \citenamefont {Cao}, \citenamefont {Stavinoha}, \citenamefont
  {Chen}, \citenamefont {Man}, \citenamefont {Carr}, \citenamefont {Singleton},
  \citenamefont {Zapf}, \citenamefont {Benavides}, \citenamefont {Chan},
  \citenamefont {Zhang}, \citenamefont {Rhodes}, \citenamefont {Chiu},
  \citenamefont {Balicas}, \citenamefont {Aczel}, \citenamefont {Huang},
  \citenamefont {Lynn}, \citenamefont {Gaudet}, \citenamefont {Sokolov},
  \citenamefont {Walker}, \citenamefont {Adroja}, \citenamefont {Dai},
  \citenamefont {Nevidomskyy}, \citenamefont {Huang},\ and\ \citenamefont
  {Morosan}}]{raiYbRh3Si7}%
  \BibitemOpen
  \bibfield  {author} {\bibinfo {author} {\bibfnamefont {B.~K.}\ \bibnamefont
  {Rai}}, \bibinfo {author} {\bibfnamefont {S.}~\bibnamefont {Chikara}},
  \bibinfo {author} {\bibfnamefont {X.}~\bibnamefont {Ding}}, \bibinfo {author}
  {\bibfnamefont {I.~W.~H.}\ \bibnamefont {Oswald}}, \bibinfo {author}
  {\bibfnamefont {R.}~\bibnamefont {Sch\"onemann}}, \bibinfo {author}
  {\bibfnamefont {V.}~\bibnamefont {Loganathan}}, \bibinfo {author}
  {\bibfnamefont {A.~M.}\ \bibnamefont {Hallas}}, \bibinfo {author}
  {\bibfnamefont {H.~B.}\ \bibnamefont {Cao}}, \bibinfo {author} {\bibfnamefont
  {M.}~\bibnamefont {Stavinoha}}, \bibinfo {author} {\bibfnamefont
  {T.}~\bibnamefont {Chen}}, \bibinfo {author} {\bibfnamefont {H.}~\bibnamefont
  {Man}}, \bibinfo {author} {\bibfnamefont {S.}~\bibnamefont {Carr}}, \bibinfo
  {author} {\bibfnamefont {J.}~\bibnamefont {Singleton}}, \bibinfo {author}
  {\bibfnamefont {V.}~\bibnamefont {Zapf}}, \bibinfo {author} {\bibfnamefont
  {K.~A.}\ \bibnamefont {Benavides}}, \bibinfo {author} {\bibfnamefont {J.~Y.}\
  \bibnamefont {Chan}}, \bibinfo {author} {\bibfnamefont {Q.~R.}\ \bibnamefont
  {Zhang}}, \bibinfo {author} {\bibfnamefont {D.}~\bibnamefont {Rhodes}},
  \bibinfo {author} {\bibfnamefont {Y.~C.}\ \bibnamefont {Chiu}}, \bibinfo
  {author} {\bibfnamefont {L.}~\bibnamefont {Balicas}}, \bibinfo {author}
  {\bibfnamefont {A.~A.}\ \bibnamefont {Aczel}}, \bibinfo {author}
  {\bibfnamefont {Q.}~\bibnamefont {Huang}}, \bibinfo {author} {\bibfnamefont
  {J.~W.}\ \bibnamefont {Lynn}}, \bibinfo {author} {\bibfnamefont
  {J.}~\bibnamefont {Gaudet}}, \bibinfo {author} {\bibfnamefont {D.~A.}\
  \bibnamefont {Sokolov}}, \bibinfo {author} {\bibfnamefont {H.~C.}\
  \bibnamefont {Walker}}, \bibinfo {author} {\bibfnamefont {D.~T.}\
  \bibnamefont {Adroja}}, \bibinfo {author} {\bibfnamefont {P.}~\bibnamefont
  {Dai}}, \bibinfo {author} {\bibfnamefont {A.~H.}\ \bibnamefont
  {Nevidomskyy}}, \bibinfo {author} {\bibfnamefont {C.-L.}\ \bibnamefont
  {Huang}}, \ and\ \bibinfo {author} {\bibfnamefont {E.}~\bibnamefont
  {Morosan}},\ }\href {\doibase 10.1103/PhysRevX.8.041047} {\bibfield
  {journal} {\bibinfo  {journal} {Phys. Rev. X}\ }\textbf {\bibinfo {volume}
  {8}},\ \bibinfo {pages} {041047} (\bibinfo {year} {2018})}\BibitemShut
  {NoStop}%
\bibitem [{\citenamefont {Nozi{\`e}re}(1985)}]{Noz1}%
  \BibitemOpen
  \bibfield  {author} {\bibinfo {author} {\bibfnamefont {P.}~\bibnamefont
  {Nozi{\`e}re}},\ }\href@noop {} {\bibfield  {journal} {\bibinfo  {journal}
  {Ann. Phys. (Paris)}\ }\textbf {\bibinfo {volume} {10}},\ \bibinfo {pages}
  {19} (\bibinfo {year} {1985})}\BibitemShut {NoStop}%
\bibitem [{\citenamefont {Nozi{\`e}re}(1998)}]{Noz2}%
  \BibitemOpen
  \bibfield  {author} {\bibinfo {author} {\bibfnamefont {P.}~\bibnamefont
  {Nozi{\`e}re}},\ }\href@noop {} {\bibfield  {journal} {\bibinfo  {journal}
  {Eur. Phys. J. B}\ }\textbf {\bibinfo {volume} {6}},\ \bibinfo {pages} {447}
  (\bibinfo {year} {1998})}\BibitemShut {NoStop}%
\bibitem [{\citenamefont {Rai}\ \emph {et~al.}(2019)\citenamefont {Rai},
  \citenamefont {Stavinoha}, \citenamefont {Banda}, \citenamefont {Hafner},
  \citenamefont {Benavides}, \citenamefont {Sokolov}, \citenamefont {Chan},
  \citenamefont {Brando}, \citenamefont {Huang},\ and\ \citenamefont
  {Morosan}}]{raiYbIr3Ge7}%
  \BibitemOpen
  \bibfield  {author} {\bibinfo {author} {\bibfnamefont {B.~K.}\ \bibnamefont
  {Rai}}, \bibinfo {author} {\bibfnamefont {M.}~\bibnamefont {Stavinoha}},
  \bibinfo {author} {\bibfnamefont {J.}~\bibnamefont {Banda}}, \bibinfo
  {author} {\bibfnamefont {D.}~\bibnamefont {Hafner}}, \bibinfo {author}
  {\bibfnamefont {K.~A.}\ \bibnamefont {Benavides}}, \bibinfo {author}
  {\bibfnamefont {D.}~\bibnamefont {Sokolov}}, \bibinfo {author} {\bibfnamefont
  {J.}~\bibnamefont {Chan}}, \bibinfo {author} {\bibfnamefont {M.}~\bibnamefont
  {Brando}}, \bibinfo {author} {\bibfnamefont {C.-L.}\ \bibnamefont {Huang}}, \
  and\ \bibinfo {author} {\bibfnamefont {E.}~\bibnamefont {Morosan}},\ }\href
  {\doibase 10.1103/PhysRevB.99.121109} {\bibfield  {journal} {\bibinfo
  {journal} {Phys. Rev. B(R)}\ }\textbf {\bibinfo {volume} {99}},\ \bibinfo
  {pages} {121109} (\bibinfo {year} {2019})}\BibitemShut {NoStop}%
\bibitem [{\citenamefont {Hundley}\ \emph {et~al.}(1990)\citenamefont
  {Hundley}, \citenamefont {Canfield}, \citenamefont {Thompson}, \citenamefont
  {Fisk},\ and\ \citenamefont {Lawrence}}]{hundley1990}%
  \BibitemOpen
  \bibfield  {author} {\bibinfo {author} {\bibfnamefont {M.}~\bibnamefont
  {Hundley}}, \bibinfo {author} {\bibfnamefont {P.}~\bibnamefont {Canfield}},
  \bibinfo {author} {\bibfnamefont {J.}~\bibnamefont {Thompson}}, \bibinfo
  {author} {\bibfnamefont {Z.}~\bibnamefont {Fisk}}, \ and\ \bibinfo {author}
  {\bibfnamefont {J.}~\bibnamefont {Lawrence}},\ }\href@noop {} {\bibfield
  {journal} {\bibinfo  {journal} {Physical Review B}\ }\textbf {\bibinfo
  {volume} {42}},\ \bibinfo {pages} {6842} (\bibinfo {year}
  {1990})}\BibitemShut {NoStop}%
\bibitem [{\citenamefont {Das}\ and\ \citenamefont
  {Sampathkumaran}(1995)}]{das1995}%
  \BibitemOpen
  \bibfield  {author} {\bibinfo {author} {\bibfnamefont {I.}~\bibnamefont
  {Das}}\ and\ \bibinfo {author} {\bibfnamefont {E.}~\bibnamefont
  {Sampathkumaran}},\ }\href@noop {} {\bibfield  {journal} {\bibinfo  {journal}
  {Physical Review B}\ }\textbf {\bibinfo {volume} {51}},\ \bibinfo {pages}
  {1308} (\bibinfo {year} {1995})}\BibitemShut {NoStop}%
\bibitem [{\citenamefont {Menth}\ \emph {et~al.}(1969)\citenamefont {Menth},
  \citenamefont {Buehler},\ and\ \citenamefont {Geballe}}]{menth1969}%
  \BibitemOpen
  \bibfield  {author} {\bibinfo {author} {\bibfnamefont {A.}~\bibnamefont
  {Menth}}, \bibinfo {author} {\bibfnamefont {E.}~\bibnamefont {Buehler}}, \
  and\ \bibinfo {author} {\bibfnamefont {T.}~\bibnamefont {Geballe}},\
  }\href@noop {} {\bibfield  {journal} {\bibinfo  {journal} {Physical Review
  Letters}\ }\textbf {\bibinfo {volume} {22}},\ \bibinfo {pages} {295}
  (\bibinfo {year} {1969})}\BibitemShut {NoStop}%
\bibitem [{\citenamefont {Kasaya}\ \emph {et~al.}(1983)\citenamefont {Kasaya},
  \citenamefont {Iga}, \citenamefont {Negishi}, \citenamefont {Nakai},\ and\
  \citenamefont {Kasuya}}]{Kasaya1983}%
  \BibitemOpen
  \bibfield  {author} {\bibinfo {author} {\bibfnamefont {M.}~\bibnamefont
  {Kasaya}}, \bibinfo {author} {\bibfnamefont {F.}~\bibnamefont {Iga}},
  \bibinfo {author} {\bibfnamefont {K.}~\bibnamefont {Negishi}}, \bibinfo
  {author} {\bibfnamefont {S.}~\bibnamefont {Nakai}}, \ and\ \bibinfo {author}
  {\bibfnamefont {T.}~\bibnamefont {Kasuya}},\ }\href@noop {} {\bibfield
  {journal} {\bibinfo  {journal} {Journal of Magnetism and Magnetic Materials}\
  }\textbf {\bibinfo {volume} {31}},\ \bibinfo {pages} {437} (\bibinfo {year}
  {1983})}\BibitemShut {NoStop}%
\bibitem [{\citenamefont {Kasaya}\ \emph {et~al.}(1985)\citenamefont {Kasaya},
  \citenamefont {Iga}, \citenamefont {Takigawa},\ and\ \citenamefont
  {Kasuya}}]{Kasaya1985}%
  \BibitemOpen
  \bibfield  {author} {\bibinfo {author} {\bibfnamefont {M.}~\bibnamefont
  {Kasaya}}, \bibinfo {author} {\bibfnamefont {F.}~\bibnamefont {Iga}},
  \bibinfo {author} {\bibfnamefont {M.}~\bibnamefont {Takigawa}}, \ and\
  \bibinfo {author} {\bibfnamefont {T.}~\bibnamefont {Kasuya}},\ }\href@noop {}
  {\bibfield  {journal} {\bibinfo  {journal} {Journal of Magnetism and Magnetic
  Materials}\ }\textbf {\bibinfo {volume} {47}},\ \bibinfo {pages} {429}
  (\bibinfo {year} {1985})}\BibitemShut {NoStop}%
\bibitem [{\citenamefont {Meisner}\ \emph {et~al.}(1985)\citenamefont
  {Meisner}, \citenamefont {Torikachvili}, \citenamefont {Yang}, \citenamefont
  {Maple},\ and\ \citenamefont {Guertin}}]{meisner1985}%
  \BibitemOpen
  \bibfield  {author} {\bibinfo {author} {\bibfnamefont {G.}~\bibnamefont
  {Meisner}}, \bibinfo {author} {\bibfnamefont {M.}~\bibnamefont
  {Torikachvili}}, \bibinfo {author} {\bibfnamefont {K.}~\bibnamefont {Yang}},
  \bibinfo {author} {\bibfnamefont {M.}~\bibnamefont {Maple}}, \ and\ \bibinfo
  {author} {\bibfnamefont {R.}~\bibnamefont {Guertin}},\ }\href@noop {}
  {\bibfield  {journal} {\bibinfo  {journal} {Journal of Applied Physics}\
  }\textbf {\bibinfo {volume} {57}},\ \bibinfo {pages} {3073} (\bibinfo {year}
  {1985})}\BibitemShut {NoStop}%
\bibitem [{\citenamefont {Syers}\ \emph {et~al.}(2015)\citenamefont {Syers},
  \citenamefont {Kim}, \citenamefont {Fuhrer},\ and\ \citenamefont
  {Paglione}}]{syers2015}%
  \BibitemOpen
  \bibfield  {author} {\bibinfo {author} {\bibfnamefont {P.}~\bibnamefont
  {Syers}}, \bibinfo {author} {\bibfnamefont {D.}~\bibnamefont {Kim}}, \bibinfo
  {author} {\bibfnamefont {M.~S.}\ \bibnamefont {Fuhrer}}, \ and\ \bibinfo
  {author} {\bibfnamefont {J.}~\bibnamefont {Paglione}},\ }\href@noop {}
  {\bibfield  {journal} {\bibinfo  {journal} {Physical review letters}\
  }\textbf {\bibinfo {volume} {114}},\ \bibinfo {pages} {096601} (\bibinfo
  {year} {2015})}\BibitemShut {NoStop}%
\bibitem [{\citenamefont {Phelan}\ \emph {et~al.}(2014)\citenamefont {Phelan},
  \citenamefont {Koohpayeh}, \citenamefont {Cottingham}, \citenamefont
  {Freeland}, \citenamefont {Leiner}, \citenamefont {Broholm},\ and\
  \citenamefont {McQueen}}]{phelan2014}%
  \BibitemOpen
  \bibfield  {author} {\bibinfo {author} {\bibfnamefont {W.}~\bibnamefont
  {Phelan}}, \bibinfo {author} {\bibfnamefont {S.}~\bibnamefont {Koohpayeh}},
  \bibinfo {author} {\bibfnamefont {P.}~\bibnamefont {Cottingham}}, \bibinfo
  {author} {\bibfnamefont {J.}~\bibnamefont {Freeland}}, \bibinfo {author}
  {\bibfnamefont {J.}~\bibnamefont {Leiner}}, \bibinfo {author} {\bibfnamefont
  {C.}~\bibnamefont {Broholm}}, \ and\ \bibinfo {author} {\bibfnamefont
  {T.}~\bibnamefont {McQueen}},\ }\href@noop {} {\bibfield  {journal} {\bibinfo
   {journal} {Physical Review X}\ }\textbf {\bibinfo {volume} {4}},\ \bibinfo
  {pages} {031012} (\bibinfo {year} {2014})}\BibitemShut {NoStop}%
\bibitem [{\citenamefont {Hafner}\ \emph {et~al.}(2019)\citenamefont {Hafner},
  \citenamefont {Rai}, \citenamefont {Banda}, \citenamefont {Kliemt},
  \citenamefont {Krellner}, \citenamefont {Sichelschmidt}, \citenamefont
  {Morosan}, \citenamefont {Geibel},\ and\ \citenamefont
  {Brando}}]{hafner2019}%
  \BibitemOpen
  \bibfield  {author} {\bibinfo {author} {\bibfnamefont {D.}~\bibnamefont
  {Hafner}}, \bibinfo {author} {\bibfnamefont {B.~K.}\ \bibnamefont {Rai}},
  \bibinfo {author} {\bibfnamefont {J.}~\bibnamefont {Banda}}, \bibinfo
  {author} {\bibfnamefont {K.}~\bibnamefont {Kliemt}}, \bibinfo {author}
  {\bibfnamefont {C.}~\bibnamefont {Krellner}}, \bibinfo {author}
  {\bibfnamefont {J.}~\bibnamefont {Sichelschmidt}}, \bibinfo {author}
  {\bibfnamefont {E.}~\bibnamefont {Morosan}}, \bibinfo {author} {\bibfnamefont
  {C.}~\bibnamefont {Geibel}}, \ and\ \bibinfo {author} {\bibfnamefont
  {M.}~\bibnamefont {Brando}},\ }\href@noop {} {\bibfield  {journal} {\bibinfo
  {journal} {Physical Review B}\ }\textbf {\bibinfo {volume} {99}},\ \bibinfo
  {pages} {201109(R)} (\bibinfo {year} {2019})}\BibitemShut {NoStop}%
\bibitem [{\citenamefont {Johansson}(1979)}]{johansson1979}%
  \BibitemOpen
  \bibfield  {author} {\bibinfo {author} {\bibfnamefont {B.}~\bibnamefont
  {Johansson}},\ }\href@noop {} {\bibfield  {journal} {\bibinfo  {journal}
  {Physical Review B}\ }\textbf {\bibinfo {volume} {19}},\ \bibinfo {pages}
  {6615} (\bibinfo {year} {1979})}\BibitemShut {NoStop}%
\bibitem [{\citenamefont {Lynn}\ \emph {et~al.}(2012)\citenamefont {Lynn},
  \citenamefont {Chen}, \citenamefont {Chang}, \citenamefont {Zhao},
  \citenamefont {Chi}, \citenamefont {Ratcliff} \emph {et~al.}}]{lynn2012}%
  \BibitemOpen
  \bibfield  {author} {\bibinfo {author} {\bibfnamefont {J.}~\bibnamefont
  {Lynn}}, \bibinfo {author} {\bibfnamefont {Y.}~\bibnamefont {Chen}}, \bibinfo
  {author} {\bibfnamefont {S.}~\bibnamefont {Chang}}, \bibinfo {author}
  {\bibfnamefont {Y.}~\bibnamefont {Zhao}}, \bibinfo {author} {\bibfnamefont
  {S.}~\bibnamefont {Chi}}, \bibinfo {author} {\bibfnamefont {W.}~\bibnamefont
  {Ratcliff}},  \emph {et~al.},\ }\href@noop {} {\bibfield  {journal} {\bibinfo
   {journal} {Journal of research of the National Institute of Standards and
  Technology}\ }\textbf {\bibinfo {volume} {117}},\ \bibinfo {pages} {61}
  (\bibinfo {year} {2012})}\BibitemShut {NoStop}%
\bibitem [{\citenamefont {Rodr{\'\i}guez-Carvajal}(1993)}]{rodriguez1993}%
  \BibitemOpen
  \bibfield  {author} {\bibinfo {author} {\bibfnamefont {J.}~\bibnamefont
  {Rodr{\'\i}guez-Carvajal}},\ }\href@noop {} {\bibfield  {journal} {\bibinfo
  {journal} {Physica B: Condensed Matter}\ }\textbf {\bibinfo {volume} {192}},\
  \bibinfo {pages} {55} (\bibinfo {year} {1993})}\BibitemShut {NoStop}%
\bibitem [{\citenamefont {Torikachvili}\ \emph {et~al.}(1986)\citenamefont
  {Torikachvili}, \citenamefont {Rossel}, \citenamefont {McElfresh},
  \citenamefont {Maple}, \citenamefont {Guertin},\ and\ \citenamefont
  {Meisner}}]{torikachvili1986}%
  \BibitemOpen
  \bibfield  {author} {\bibinfo {author} {\bibfnamefont {M.}~\bibnamefont
  {Torikachvili}}, \bibinfo {author} {\bibfnamefont {C.}~\bibnamefont
  {Rossel}}, \bibinfo {author} {\bibfnamefont {M.}~\bibnamefont {McElfresh}},
  \bibinfo {author} {\bibfnamefont {M.}~\bibnamefont {Maple}}, \bibinfo
  {author} {\bibfnamefont {R.}~\bibnamefont {Guertin}}, \ and\ \bibinfo
  {author} {\bibfnamefont {G.}~\bibnamefont {Meisner}},\ }\href@noop {}
  {\bibfield  {journal} {\bibinfo  {journal} {journal of Magnetism and Magnetic
  Materials}\ }\textbf {\bibinfo {volume} {54}},\ \bibinfo {pages} {365}
  (\bibinfo {year} {1986})}\BibitemShut {NoStop}%
\bibitem [{\citenamefont {Cooley}\ \emph {et~al.}(1999)\citenamefont {Cooley},
  \citenamefont {Mielke}, \citenamefont {Hults}, \citenamefont {Goettee},
  \citenamefont {Honold}, \citenamefont {Modler}, \citenamefont {Lacerda},
  \citenamefont {Rickel},\ and\ \citenamefont {Smith}}]{cooley1999}%
  \BibitemOpen
  \bibfield  {author} {\bibinfo {author} {\bibfnamefont {J.}~\bibnamefont
  {Cooley}}, \bibinfo {author} {\bibfnamefont {C.}~\bibnamefont {Mielke}},
  \bibinfo {author} {\bibfnamefont {W.}~\bibnamefont {Hults}}, \bibinfo
  {author} {\bibfnamefont {J.}~\bibnamefont {Goettee}}, \bibinfo {author}
  {\bibfnamefont {M.}~\bibnamefont {Honold}}, \bibinfo {author} {\bibfnamefont
  {R.}~\bibnamefont {Modler}}, \bibinfo {author} {\bibfnamefont
  {A.}~\bibnamefont {Lacerda}}, \bibinfo {author} {\bibfnamefont
  {D.}~\bibnamefont {Rickel}}, \ and\ \bibinfo {author} {\bibfnamefont
  {J.}~\bibnamefont {Smith}},\ }\href@noop {} {\bibfield  {journal} {\bibinfo
  {journal} {Journal of superconductivity}\ }\textbf {\bibinfo {volume} {12}},\
  \bibinfo {pages} {171} (\bibinfo {year} {1999})}\BibitemShut {NoStop}%
\bibitem [{\citenamefont {Perdew}\ \emph {et~al.}(1996)\citenamefont {Perdew},
  \citenamefont {Burke},\ and\ \citenamefont {Ernzerhof}}]{perdew1996}%
  \BibitemOpen
  \bibfield  {author} {\bibinfo {author} {\bibfnamefont {J.~P.}\ \bibnamefont
  {Perdew}}, \bibinfo {author} {\bibfnamefont {K.}~\bibnamefont {Burke}}, \
  and\ \bibinfo {author} {\bibfnamefont {M.}~\bibnamefont {Ernzerhof}},\
  }\href@noop {} {\bibfield  {journal} {\bibinfo  {journal} {Physical review
  letters}\ }\textbf {\bibinfo {volume} {77}},\ \bibinfo {pages} {3865}
  (\bibinfo {year} {1996})}\BibitemShut {NoStop}%
\bibitem [{\citenamefont {Blaha}\ \emph {et~al.}(2001)\citenamefont {Blaha},
  \citenamefont {Schwarz}, \citenamefont {Madsen}, \citenamefont {Kvasnicka},\
  and\ \citenamefont {Luitz}}]{blaha2001}%
  \BibitemOpen
  \bibfield  {author} {\bibinfo {author} {\bibfnamefont {P.}~\bibnamefont
  {Blaha}}, \bibinfo {author} {\bibfnamefont {K.}~\bibnamefont {Schwarz}},
  \bibinfo {author} {\bibfnamefont {G.~K.}\ \bibnamefont {Madsen}}, \bibinfo
  {author} {\bibfnamefont {D.}~\bibnamefont {Kvasnicka}}, \ and\ \bibinfo
  {author} {\bibfnamefont {J.}~\bibnamefont {Luitz}},\ }\href@noop {}
  {\bibfield  {journal} {\bibinfo  {journal} {An augmented plane wave+ local
  orbitals program for calculating crystal properties}\ } (\bibinfo {year}
  {2001})}\BibitemShut {NoStop}%
\bibitem [{\citenamefont {Anisimov}\ \emph {et~al.}(1993)\citenamefont
  {Anisimov}, \citenamefont {Solovyev}, \citenamefont {Korotin}, \citenamefont
  {Czy{\.z}yk},\ and\ \citenamefont {Sawatzky}}]{anisimov1993}%
  \BibitemOpen
  \bibfield  {author} {\bibinfo {author} {\bibfnamefont {V.~I.}\ \bibnamefont
  {Anisimov}}, \bibinfo {author} {\bibfnamefont {I.}~\bibnamefont {Solovyev}},
  \bibinfo {author} {\bibfnamefont {M.}~\bibnamefont {Korotin}}, \bibinfo
  {author} {\bibfnamefont {M.}~\bibnamefont {Czy{\.z}yk}}, \ and\ \bibinfo
  {author} {\bibfnamefont {G.}~\bibnamefont {Sawatzky}},\ }\href@noop {}
  {\bibfield  {journal} {\bibinfo  {journal} {Physical Review B}\ }\textbf
  {\bibinfo {volume} {48}},\ \bibinfo {pages} {16929} (\bibinfo {year}
  {1993})}\BibitemShut {NoStop}%
\end{thebibliography}%

\newpage
\section*{Supplementary Materials}
\beginsupplement

\subsection{Crystallography}\label{App.cryst}

Single crystals of YbIr$_3$Si$_7$ were oriented using back scattering x-ray Laue diffraction along [100] and [001] in the equivalent hexagonal unit cell setting. Each crystal was mounted onto a three-axis goniometer, and the goniometer was placed onto the stage of a MWL101 real-time back-reflection Laue camera system from Multiwire Laboratories, Ltd. All images were collected using an applied voltage of 10 kV and current of 10 mA to the tungsten x-ray source. Total counting times for all images varied but did not exceed five minutes.

\renewcommand{\arraystretch}{1}
\setlength{\tabcolsep}{40pt}

\begin{table*}[htbp]
\caption{\label{TableS1} Crystallographic parameters of YbIr$_3$Si$_7$ single crystals at $T$ = 173 K ($R\bar{3}c$)}
\begin{tabular}{c|c}
 \hline
	Formula  									&  YbIr$_3$Si$_7$	          \\ 
	$a$ (\AA) 									&  7.5562(3)   	          \\
	$c$ (\AA)									& 20.0662(11)                   \\	
	$V$ (\AA$ ^{3}$) 								& 992.21(9) 	                     \\
          Crystal dimensions (mm$^3$)                                                            & 0.1 x 0.07 x 0.07           \\
         $\theta$ range (\degree)                                                                   & 3.7 - 30.6                       \\
          Extinction coefficient                                                                          & 0.00072(2)                      \\
	absorption coefficient (mm$^{-1}$)					& 75.329                 	            \\		
	measured reflections							& 7570 		 	 \\
	independent reflections 							& 377			            \\
	R$_{int}$	 								& 0.0807		            \\
	goodness-of-fit on F$^2$	  						& 1.053		            \\
	$R_1(F)$ for ${F^2}_o \textgreater 2\sigma ({F^2}_o)^a$	& 0.0220	                       \\
	$wR_2({F^2}_o)^b$							& 0.0379                              \\ \hline

 \end{tabular}
$^{a}R_1 = \sum\mid\mid F_o\mid - \mid F_c\mid \mid / \sum \mid F_o \mid~~~^bwR_2 = [\sum[w({F_o}^2 - {F_c}^2)^2]/ \sum[w({F_o}^2)^2]]^{1/2}$ 
\end{table*}

The rhombohedral $R\bar{3}c$ space group was confirmed in YbIr$_3$Si$_7$ and LuIr$_3$Si$_7$ using powder x-ray diffraction refinement. Additionally, single crystal x-ray diffraction refinement was used to confirm the $R\bar{3}c$ space group and full occupancies of the crystallographic sites in YbIr$_3$Si$_7$. Powder x-ray diffraction measurements were performed at ambient temperature using a Bruker D8 Advance diffractometer with Cu K$\alpha$ radiation. Rietveld refinement was done using the TOPAS software package. Single crystal x-ray diffraction was conducted using a Rigaku SCX Mini diffractometer with Mo K$\alpha$ radiation. Integration of raw frame data was done using CRYSTALCLEAR 2.0. Refinement of the diffraction data was performed using XPREP and SHELXTL software packages. 

\begin{figure*}[t!]
	\includegraphics[width=2\columnwidth]{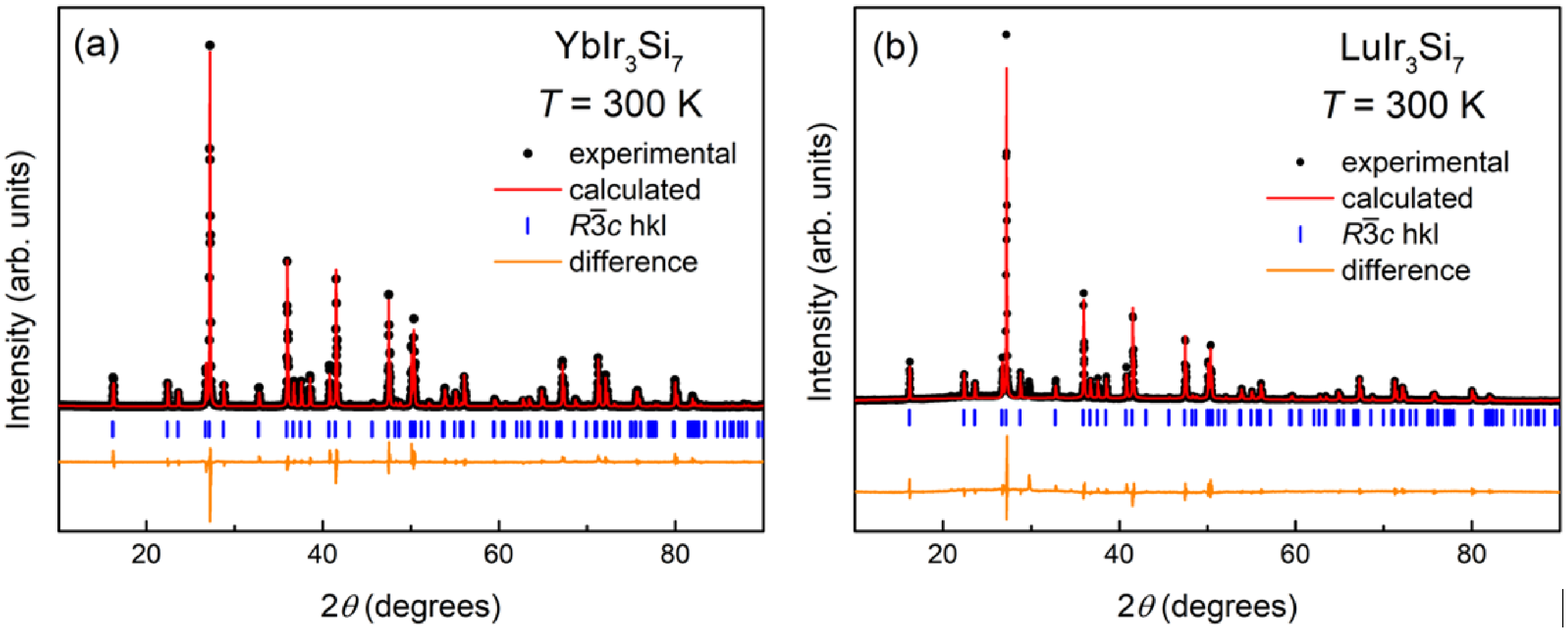}
	\caption{\label{FigS1} Powder x-ray diffraction and refinement of YbIr$_3$Si$_7$ and LuIr$_3$Si$_7$.}
\end{figure*}

Electron probe microanalysis (EPMA) was performed to obtain quantitative analysis of the Yb-Ir-Si and Lu-Ir-Si phases using wavelength dispersive spectrometry (WDS). The compositions were shown to be Yb$_{0.99(1)}$Ir$_{3.00(2)}$Si$_{7.01(2)}$ and Lu$_{1.02(3)}$Ir$_{3.04(4)}$Si$_{6.95(6)}$. The data was acquired at the EPMA laboratory, Earth Science Department, Rice University, using a JEOL JXA 8530F Hyperprobe equipped with a field emission (Schottky) emitter and five WDS spectrometers. The analytical conditions used were 15 kV accelerating voltage, 50 nA beam current, 300 nm beam diameter, and counting times of 10 seconds per peak and 5 seconds per each lower and upper background, respectively. The standards used for element calibration were synthetic metals (Si metal and Ir metal) for Si and Ir, respectively, and REE2 glass8 for Lu and Yb. Both Ir and Yb were concomitantly analyzed on two different spectrometers using two types of diffracting crystals (LiF and LiFH for Ir and PET and PETL for Yb) in order to improve the x-ray statistics and increase the accuracy of the analysis. LiFH and PETL diffracting crystals have much higher sensibility (3-4 times higher x-ray intensities) compared to LiF and PET crystals. The PhiRoZ matrix correction was employed for quantification.  

\subsection{Thermodynamic and Transport Properties}\label{App.cryst}

Resistivity and specific heat measurements were performed in a Quantum Design (QD) physical property measurement system (PPMS) equipped with a $^3$He insert. Temperature-dependent ac resistivity measurements were performed using $i$ = 0.05 mA ($i$ = 0.1 mA above 50 K) and $f$ = 35.54 Hz for a duration of 3 s. Resistivity measurements performed as a function of sample thickness were conducted on the same sample thinned using mechanical polishing with the leads reapplied before each measurement. Resistivity measurements performed up to 14 T were done using a QD DynaCool physical properties measurement system equipped with a dilution refrigerator. The electrical transport option for resistivity was performed using $i$ = 0.1 - 1 mA, $f$ = 18.3 Hz, and an averaging time of 2.3 s. Specific heat measurements were performed using an adiabatic relaxation method. A QD magnetic properties measurement system was used to perform dc magnetic susceptibility measurements up to 400 K. High-temperature magnetic susceptibility measurements were carried out on a QD PPMS equipped with the VSM oven option. The measurements covered a temperature range from 300 K to 1000 K. The system calibration was confirmed both before and after the measurements using a palladium calibration standard, the system showing a repeatability of better than 0.2\%. Each sample was measured at least three times with each run agreeing within the noise floor of the system.  

For the neutron diffraction measurements, the sample was loaded into a liquid helium cryostat with a base temperature of 1.5 K.  To search for magnetic scattering the high intensity/coarse resolution BT-7 spectrometer was employed in two-axis mode, with a fixed initial neutron energy of 14.7 meV (wavelength 2.359 \AA) and collimator (full-width-half-maximum) configuration open—PG(002) monochromator—80'—sample—80' radial-collimator—position-sensitive detector.\cite{lynn2012} The refinements were determined by the Rietveld method with the FullProf suite.\cite{rodriguez1993}

\begin{figure*}[t!]
	\includegraphics[width=2\columnwidth]{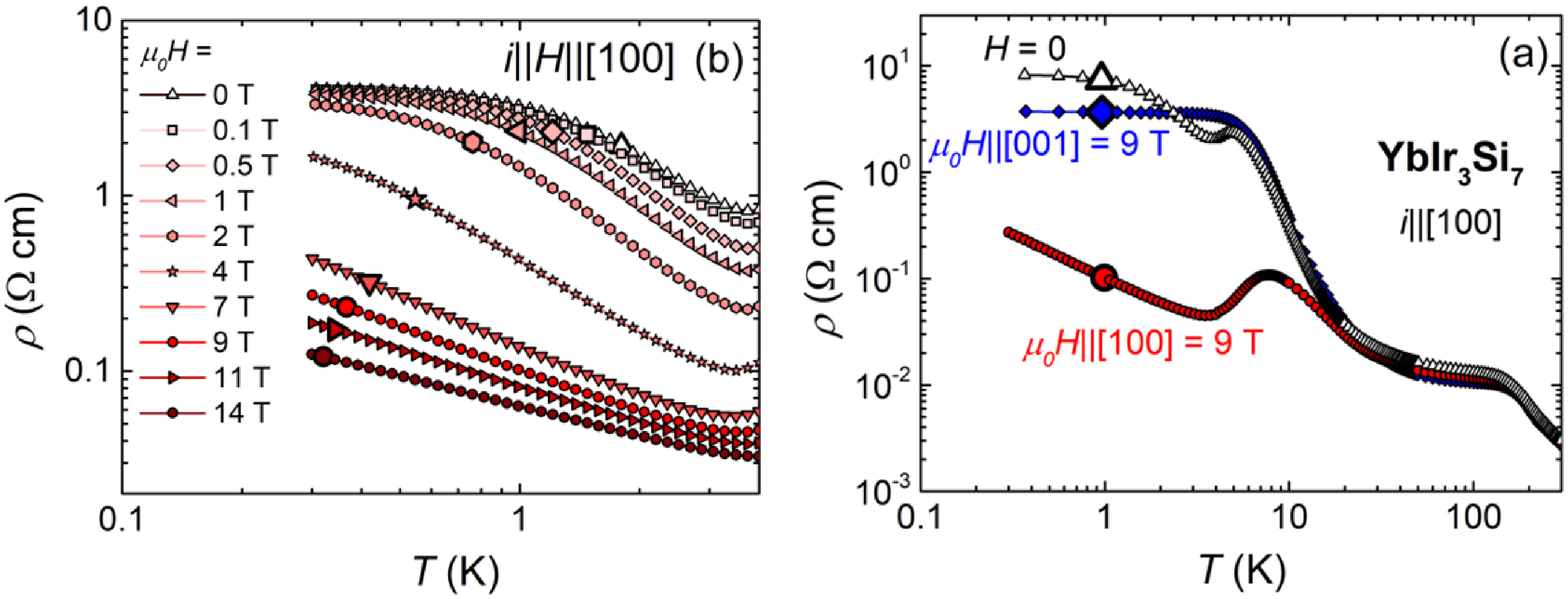}
	\caption{\label{FigS2} (a) Log-log plot of the temperature-dependent resistivity in $H$ = 0 (triangles) and in applied magnetic field $\mu_0H$ = 9 T, $H$$\parallel$[100] (circles) and $H$$\parallel$[001] (diamonds), and (b) $H$$\parallel$[100] from 0 to 14 T.}
\end{figure*}

Fig. \ref{FigS2} shows details of the electrical resistivity $\rho$($T$) in a log-log scale, for (a) $\mu_0H$ = 0 (triangles), 9 T and $H$$\parallel$[100] (circles) and $H$$\parallel$[001] (diamonds), and (b) $i$$\parallel$$H$$\parallel$[100], with applied field values up to 14 T. The preliminary anisotropic $\rho$($T$) data (Fig. \ref{FigS2}(a)) points to a qualitative change of the temperature dependence with increasing field. Indeed, the applied magnetic field (Fig. \ref{FigS2}(b)) changes the temperature dependence from - ln$T$ ($H$ = 0, open triangles) to power law -$T$$^\alpha$ above $\mu_0H$ $\sim$ 7 T. In a Kondo screening scenario, it can indeed be expected that the ln$T$ divergence of the resistivity at low $T$ changes to a power law with band polarization in applied magnetic field. 

\begin{figure}[b!]
	\includegraphics[width=1\columnwidth]{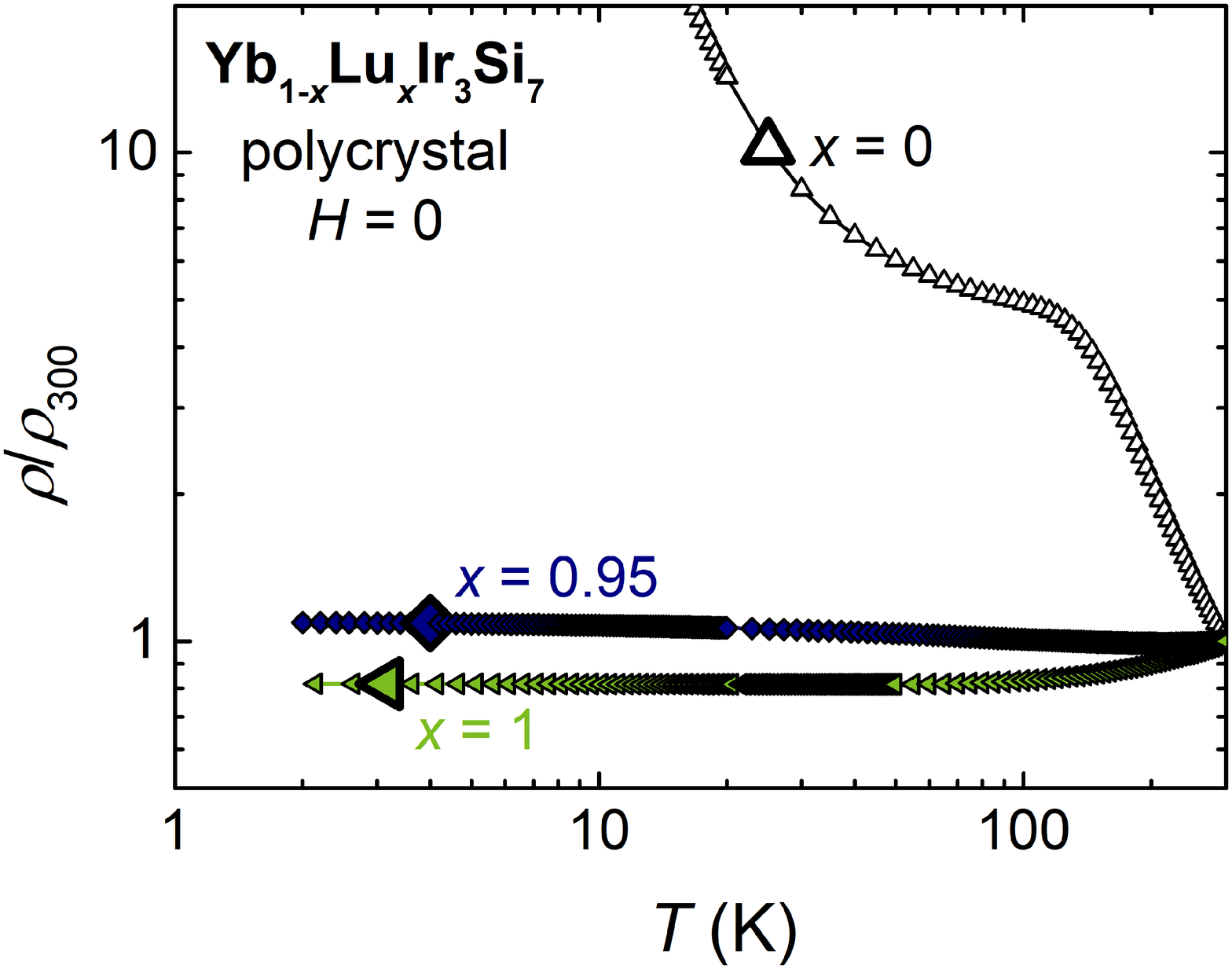}
	\caption{\label{FigS3} Scaled resistivity measurements of the doped Yb$_{1-x}$Lu$_x$Ir$_3$Si$_7$ polycrystalline samples for $x$ = 0 (open triangles), $x$ = 0.95 (diamonds) and $x$ = 1 (full triangles).}
\end{figure}

\begin{figure}[b!]
	\includegraphics[width=1\columnwidth]{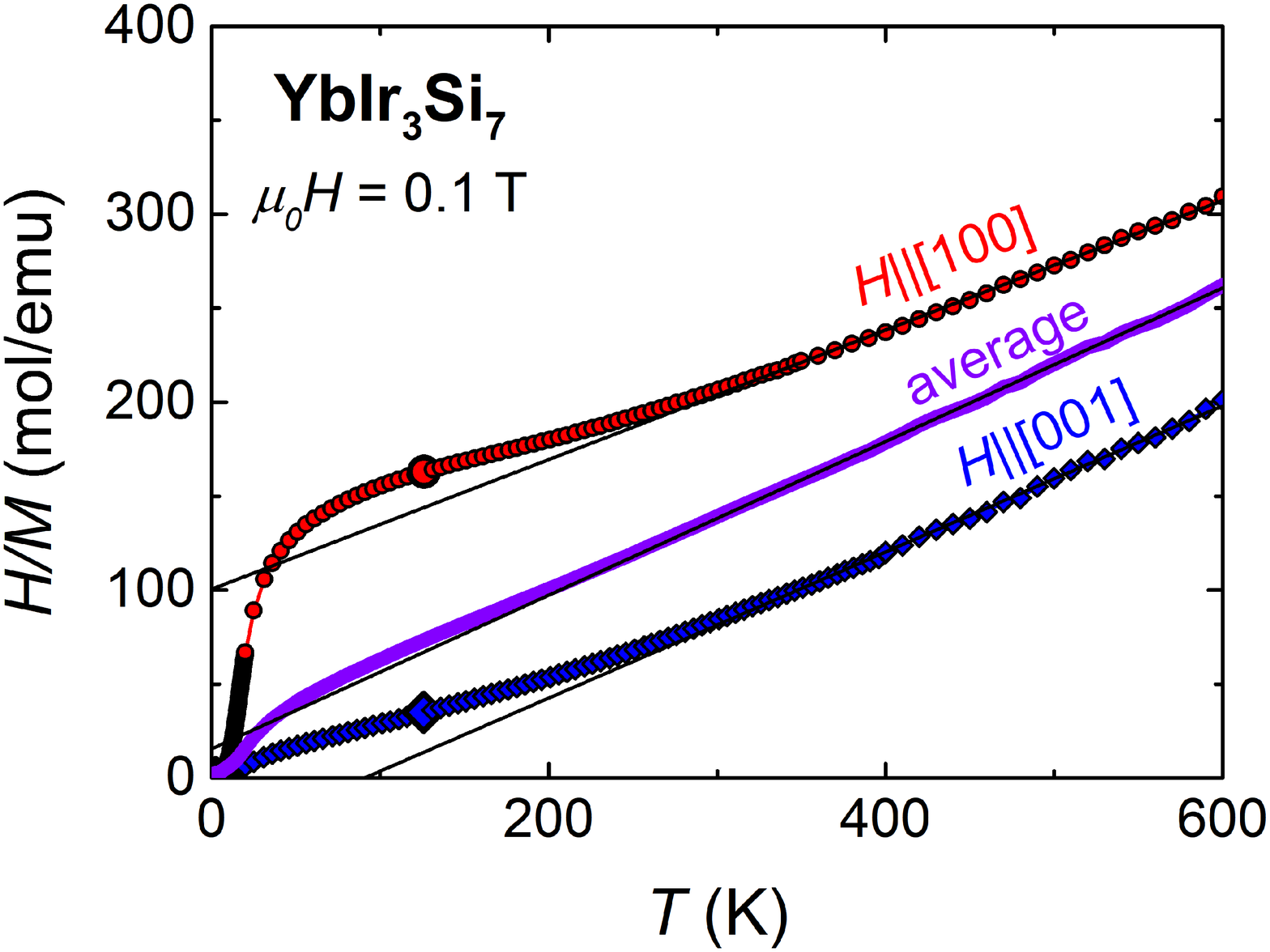}
	\caption{\label{FigS4} Anisotropic inverse magnetic susceptibility measurements (symbols) and calculated polycrystalline average $H/M$$_{ave}$ (line) with $M$$_{ave}$ = 2/3$M_{100}$ + 1/3$M_{001}$.  Solid black lines are linear fits above 350 K, evidence of Curie-Weiss behavior.}
\end{figure}

Non-magnetic Lu doping in YbIr$_3$Si$_7$ was used in an attempt to characterize the $\sim$ 100 K inflection in the $H$ = 0 $\rho$($T$) data. Single-ion CEF effects would be expected to scale with the number of magnetic Yb ions (1-$x$). Since the resistivity inflection has disappeared in the highly diluted $x$ = 0.95 sample (Fig. \ref{FigS3}, diamonds), it is unlikely due to CEF effects, rendering the onset of Kondo correlations the most likely scenario. Indeed, anisotropic inverse magnetic susceptibility data shown in Fig. \ref{FigS4} indicates that CEF splitting occurs at much higher temperatures since no deviations from the Curie-Weiss are observed in the average inverse magnetic susceptibility (solid line, Fig. \ref{FigS4}) up to 600 K.

\begin{figure*}[t!]
	\includegraphics[width=2\columnwidth]{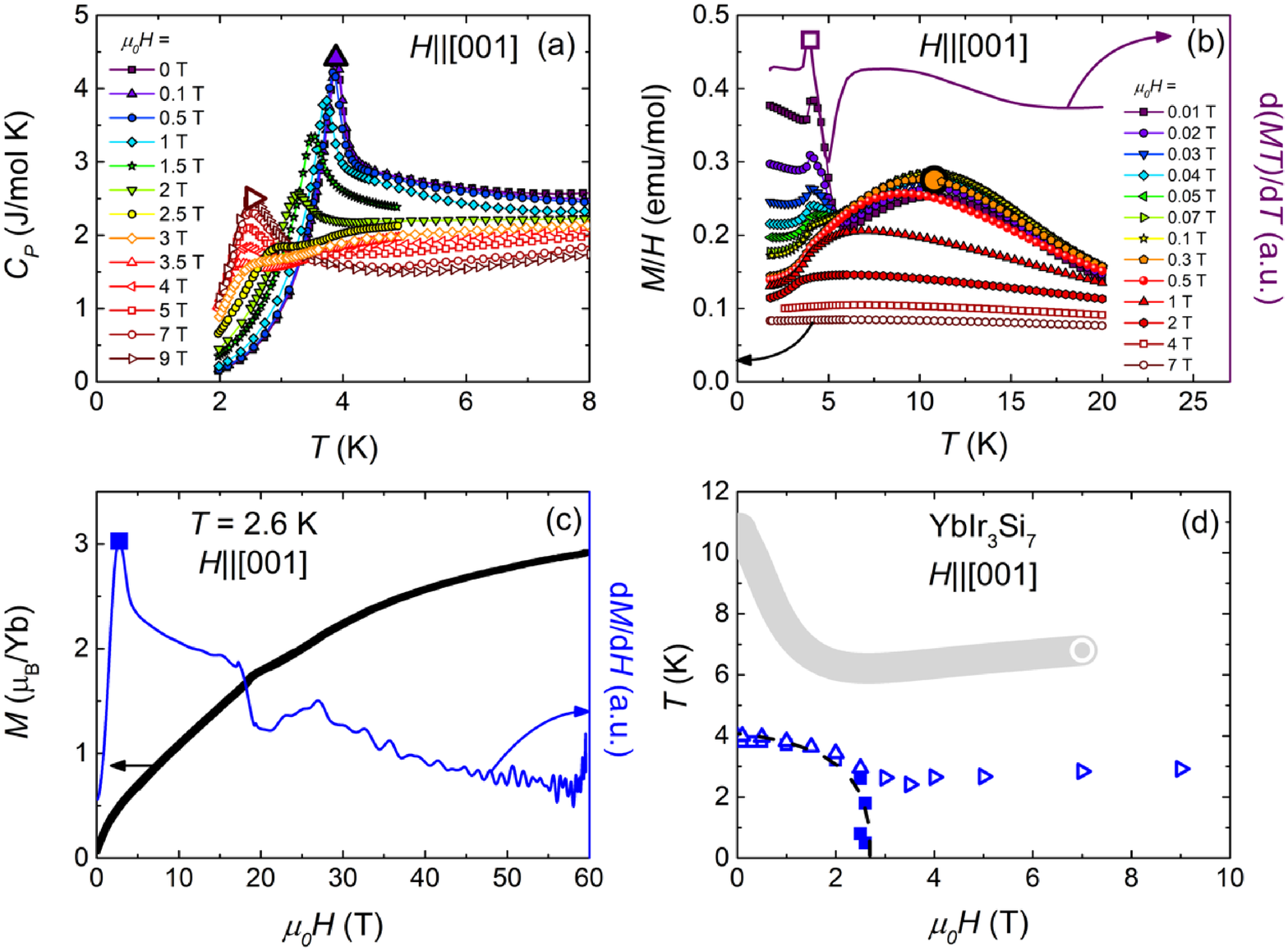}
	\caption{\label{FigS5} Temperature-dependent (a) specific heat $C_P$ and (b) magnetic susceptibility $M$/$H$ measurements in applied magnetic fields up to 14 T, for $H$$\parallel$[001]. (c) $T$ = 2.6 K magnetization isotherm $M$($H$) (thick black line, left axis) and the derivative $dM$/$dH$ (thin blue line, right axis), with the large square marking the low-$H$ metamagnetic transition. (d) $T$ – $H$ phase diagram for $H$$\parallel$[001]. The phase diagram is constructed using $C_P$ measurements, where the upward triangle represents the peak from $\mu_0H$ = 0 – 2.5 T, and the right-facing triangle represents the peak from $\mu_0H$ = 3 – 9 T.  Squares represent the peak position in $d$($MT$)/$dT$ (open symbols) and $dM$/$dH$ (full symbols). The grey line represent the broad peak observed in $M$/$H$ measurements with $H$$\parallel$[001], likely a cross-over.}
\end{figure*}

Fig. \ref{FigS5} shows the temperature-dependent specific heat $C_P$ (a) and magnetic susceptibility $M$/$H$ (b) measured in different fields, as well as an example of an $M$($H$) isotherm (c, $T$ = 2.6 K), for $H$$\parallel$[001]. Increasing field moves the antiferromagnetic transition to lower temperatures $T \rm_N$ (full triangle). However, above $\mu_0H$ = 2.5 T, the $C_P$ peak defines a new phase boundary, as $T$ moves to slightly larger values as $\mu_0H$ increase up to 9 T (open side triangle). The AFM phase boundary is also marked by a peak in $M$($T$) (symbols, b), with $T \rm_N$ determined from the peak in $d$($MT$)/$dT$ (open square) and the metamagnetic transition (Fig. \ref{FigS5}(c)) best evidenced as the maximum (full square) in the derivative $dM$/$dH$ (line). The resulting phase diagram is shown in Fig. \ref{FigS5}(d). A as-of-yet unexplained broad maximum in $M$($T$) (Fig. \ref{FigS5}(b), large circle) is depicted as a cross-over (grey line) on the phase diagram. 

\begin{figure}[b!]
	\includegraphics[width=1\columnwidth]{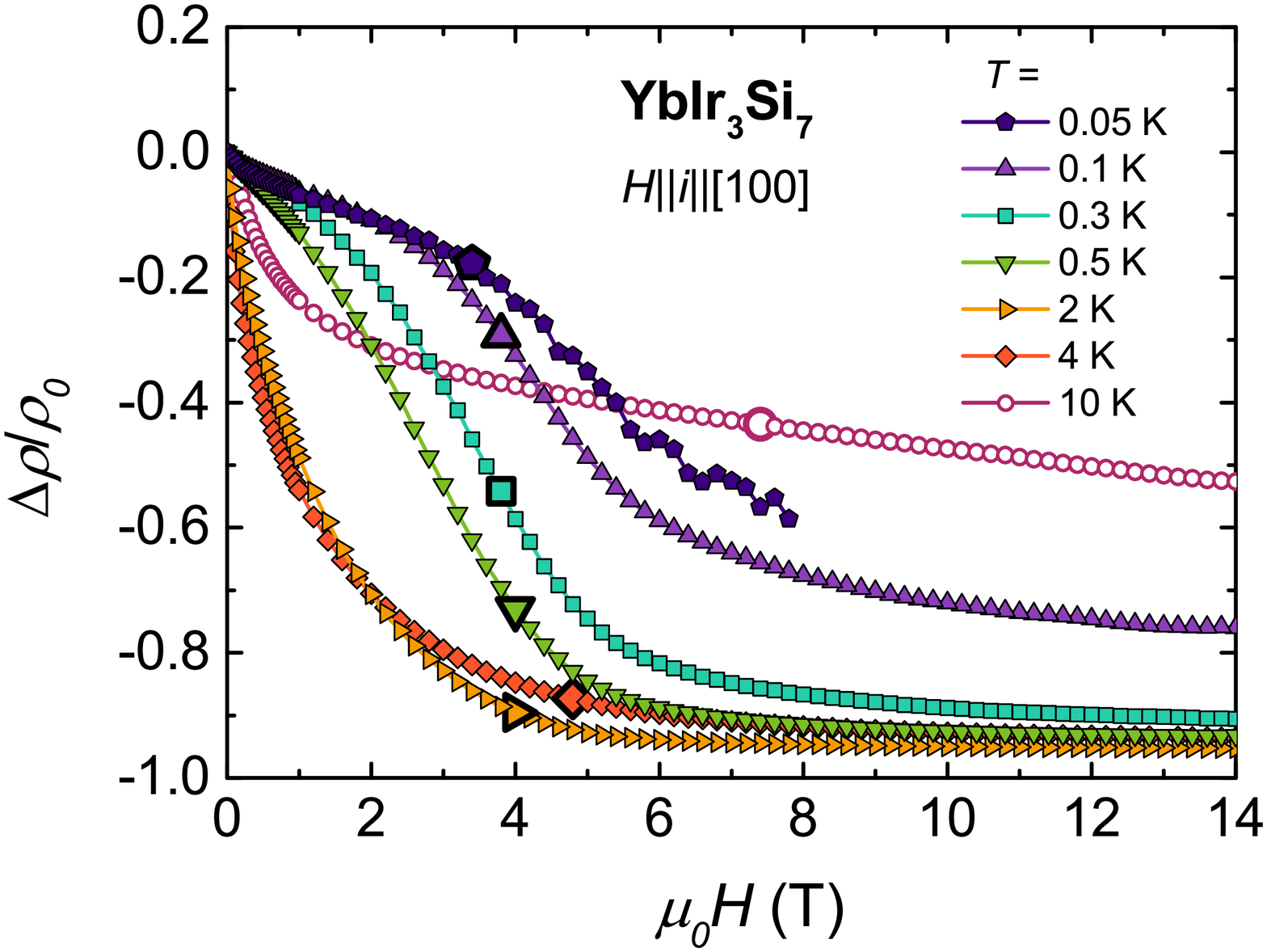}
	\caption{\label{FigS6} Longitudinal magnetoresistance $i$$\parallel$$H$$\parallel$[100] below (full symbols) and above (open symbols) $T \rm_N$.}
\end{figure}

The complex field dependence of the thermodynamic measurements motivated an investigation of the field dependence of the magnetoresistance. Large longitudinal magnetoresistance (MR) is registered in YbIr$_3$Si$_7$ (Fig. \ref{FigS6}). At $T$ = 10 K ($T$ $>$ $T \rm_N$, open circles), a moderately large MR(14T) $\sim$ -53\% was measured with $i$$\parallel$$H$$\parallel$[100]. For temperatures below $T \rm_N$ (full symbols), the negative MR increases to a maximum of $\sim$ -95\% and plateaus above 8 T. This large, negative MR below the ordering temperature is similar to that of the magnetic KI UFe$_4$P$_{12}$, where the MR approaches -90\% near the ordering temperature,\cite{torikachvili1986} attributed to the delocalization of the conduction electrons by disturbing the Kondo screening with applied magnetic field. SmB$_6$ reaches a similarly large, negative MR, albeit at a much higher applied field of 86 T when the hybridization gap is closed.\cite{cooley1999} This field value is nearly an order of magnitude larger than in YbIr$_3$Si$_7$, consistent with a Kondo temperature of $\sim$ 80 K in the former,\cite{phelan2014} an order of magnitude larger than $T \rm_K$ in the present compound.

\subsection{Band Structure Calculations}\label{App.cryst}

\begin{figure*}[t!]
	\includegraphics[width=2\columnwidth]{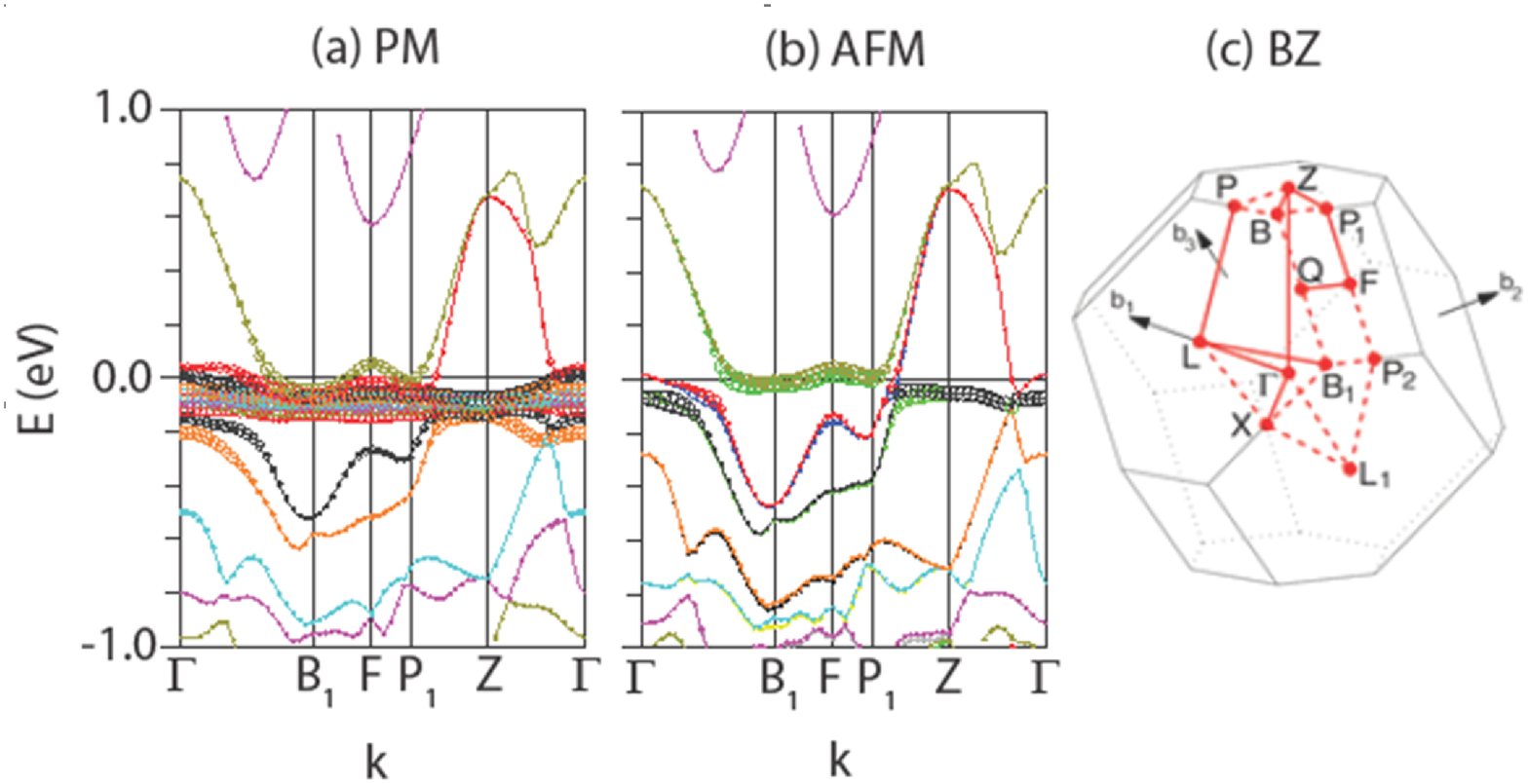}
	\caption{\label{FigS7} First principles DFT band structure of YbIr$_3$Si$_7$ in (a) PM and (b) AFM states plotted along the high-symmetry lines in the Brillouin zone shown in panel (c). The results in panel (b) were obtained within the DFT+U method ($U$ = 4 eV) to approximately capture the effect of electron correlations on Yb ion.}
\end{figure*}

In order to obtain an insight into the electronic properties of YbIr$_3$Si$_7$, we have performed first principles density functional theory (DFT) calculations using the generalized gradient approximation for the exchange-correlation functional following Perdew, Burke and Ernzerhof (PBE).\cite{perdew1996} The full-potential linearized augmented plane wave (LAPW) method was used, as implemented in the Wien2k package,\cite{blaha2001} with the scalar spin-orbit coupling included in the calculation. We have performed the electronic structure in the paramagnetic (PM) and the antiferromagnetically (AFM) ordered ground state of YbIr$_3$Si$_7$, as determined from the neutron scattering (see Fig. \ref{Neutron} in the main text). The resulting band structure in the PM state is shown in Fig. \ref{FigS7}(a), plotted along the high-symmetry lines in the Brillouin zone (BZ), with the radius of the circles reflecting the $f$ electron contribution to the respective Bloch band. In an attempt to capture the effect of strong correlations on the $f$-orbitals, we have performed the DFT+U calculations\cite{anisimov1993} in the AFM state, with the Hubbard $U$ = 4 eV on Yb ions. This results in one $f$-band at the chemical potential, shown in Fig. \ref{FigS7}(b), which is split from the other six $f$-bands that now lie about 4 eV below (not shown). 

As Fig. \ref{FigS7} shows, the nearly flat Yb 4$f$ band (light green in the center of the plot) hybridizes with the dispersive conduction electron band. Intriguingly, there is another conduction band (inverted parabola centered around the Z point, shown in red in Fig. \ref{FigS7}(a), (b)), which appears not to hybridize with the $f$ band. As a result, there is a non-vanishing electron density of the states at the chemical potential in both the PM and AFM states. Based on these results, DFT would predict YbIr$_3$Si$_7$ to be a metal, contrary to the insulating behavior of the bulk resistivity reported in Fig. \ref{Transport} in the main text. The Kondo hybridization, central to understanding this behavior, is a many-body effect not captured by the single-particle DFT calculations. Other, non-perturbative techniques are desirable in order to understand this effect. However, they are beyond the scope of the present work and are left for an upcoming study. 

\subsection{Yb Valence Determination Using XPS}\label{App.cryst}
\begin{figure}[b!]
	\includegraphics[width=1\columnwidth]{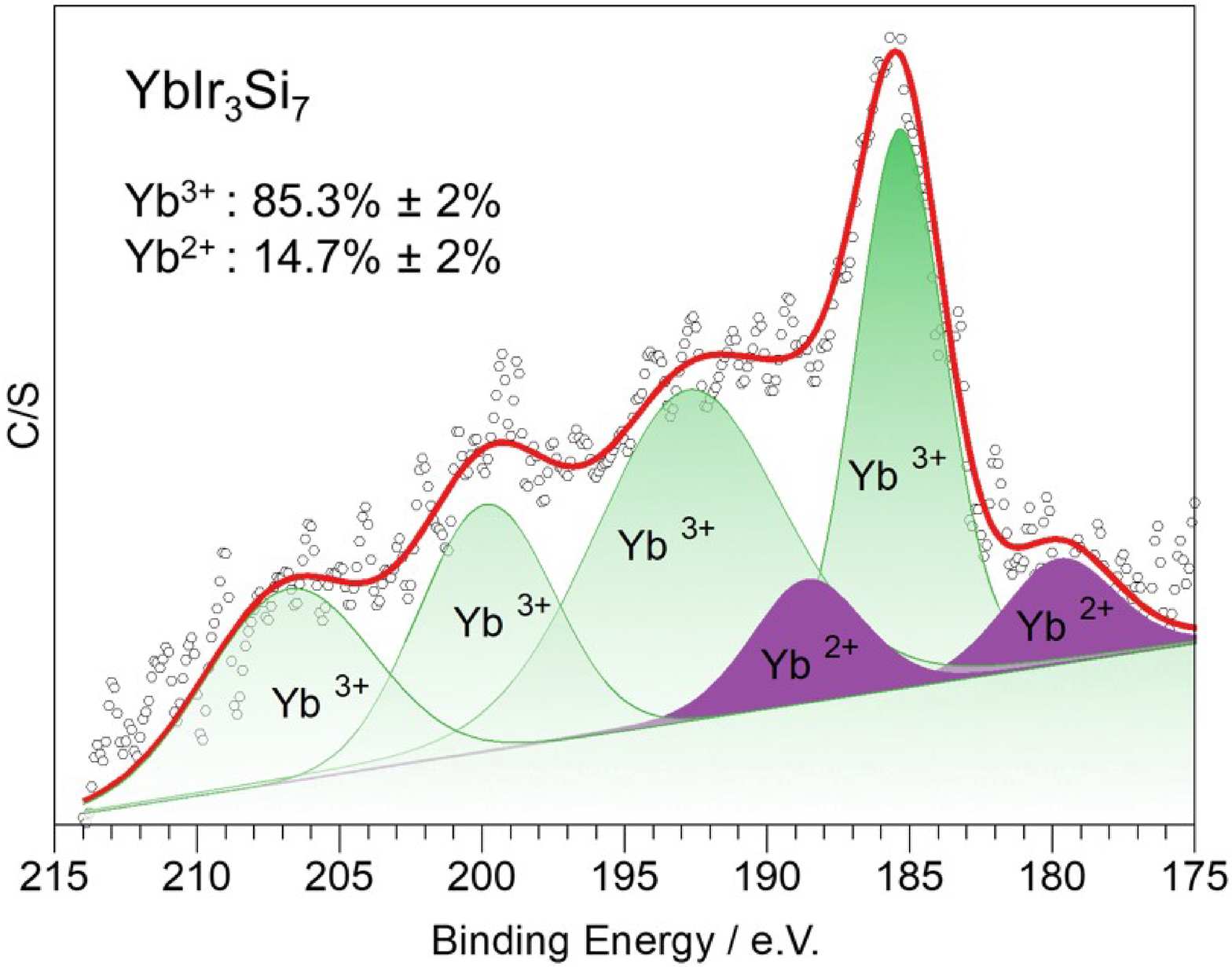}
	\caption{\label{FigS8} XPS measurements on one point of YbIr$_3$Si$_7$ indicate the ratio of Yb$^{3+}$ to Yb$^{2+}$ on the surface of a single crystal.}
\end{figure}

XPS was used to determine the ratio of Yb$^{3+}$ to Yb$^{2+}$ in YbIr$_3$Si$_7$ by measuring several spots approximately 7.5 microns in diameter and 10 nm deep. A PHI Quantera II XPS spectrometer with monochromated Al K$\alpha$ at 1486.6 eV and 26 eV pass energy was used to obtain the XPS spectra. We measured several points, and the Yb 4$d$ spectrum at one point is shown in Fig. \ref{FigS8}. It is very complex and constructed of six peaks that correspond either to Yb$^{3+}$ or Yb$^{2+}$ states. The relative abundance of Yb$^{3+}$ vs. Yb$^{2+}$ is computed from the area under each peak. For the measured volume close to the crystal surface, Yb appears to be in a mixed valence, with an average ratio Yb$^{3+}$ : Yb$^{2+}$ = 85.3 : 14.7, and this composition is consistent between all measured points. Given that bulk measurements (neutron diffraction and magnetization) only detect Yb$^{3+}$ valence (when the volume to surface ratio is much larger than in the XPS measurements), this mixed valence appears likely due to Yb$^{2+}$ valence predominantly on the surface.

\end{document}